\documentclass[12pt]{article}
\usepackage{amsmath,amssymb,amsfonts,color,graphicx,cite,soul}
\usepackage{colordvi}
\usepackage{subfigure}
\input{paperdef}
 
\graphicspath{{figs/}}

\evensidemargin \oddsidemargin
\allowdisplaybreaks

\begin{document}
\thispagestyle{empty}

\def\thefootnote{\fnsymbol{footnote}}

\begin{flushright}
MPP--2015--97\\ 
ZU-TH 11/15%\\
%arXiv:1503.nnnn [hep-ph]
\end{flushright}

\vspace{1cm}

\begin{center}

{\Large \bf{Renormalization scheme dependence of the\\%[.4em] 
two-loop QCD corrections to the neutral \\[.4em]
Higgs-boson masses in the MSSM}}

\vspace{1cm}

{\sc
S.~Borowka$^{1}$%
\footnote{email: sborowka@physik.uzh.ch}%
, T.~Hahn$^{2}$%
\footnote{email: hahn@mpp.mpg.de}%
, S.~Heinemeyer$^{3}$%
\footnote{email: Sven.Heinemeyer@cern.ch}%
, G.~Heinrich$^{2}$%
\footnote{email: gudrun@mpp.mpg.de}%
~and W.~Hollik$^{2}$%
\footnote{email: hollik@mpp.mpg.de}
}

\vspace*{.7cm}

{\sl
$^1$Institute for Physics, University of Zurich, Winterthurerstr.190,\\
8057 Zurich, Switzerland

\vspace*{0.1cm}

$^2$Max-Planck-Institut f\"ur Physik (Werner-Heisenberg-Institut),\\
F\"ohringer Ring 6, D--80805 M\"unchen, Germany

\vspace*{0.1cm}

$^3$Instituto de F\'isica de Cantabria (CSIC-UC), Santander, Spain
}

\end{center}

\vspace*{0.1cm}

\begin{abstract}
\noindent
Reaching a theoretical accuracy in the prediction of the lightest MSSM 
Higgs-boson mass, $\Mh$, at the level of the current experimental 
precision requires the inclusion of momentum-dependent contributions at 
the two-loop level. Recently two groups presented the two-loop QCD 
momentum-dependent corrections to $\Mh$~\cite{Mh-p2-BH4,Mh-p2-DDVS}, 
using a hybrid on-shell--\DRbar\ scheme, with apparently different 
results.  We show that the differences can be traced back to a different 
renormalization of the top-quark mass, and that the claim in 
\citere{Mh-p2-DDVS} of an inconsistency in \citere{Mh-p2-BH4} is 
incorrect.  We furthermore compare consistently the results for $\Mh$ 
obtained with the top-quark mass renormalized on-shell and \DRbar.  The 
latter calculation has been added to the \fh\ package and can be 
used to estimate missing higher-order corrections beyond the two-loop 
level.

%genuine scheme dependence of OS vs DRbar
\end{abstract}
%\pacs{}

\def\thefootnote{\arabic{footnote}}
\setcounter{page}{0}
\setcounter{footnote}{0}

\newpage

%%%%%%%%%%%%%%%%%%%%%%%%%%%%%%%%%%%%%%%%%%%%%%%%%%%%%%%%%%%%%%%%%%%%%%%%%%%%%%%
%%%%%%%%%%%%%%%%%%%%%%%%%%%%%%%%%%%%%%%%%%%%%%%%%%%%%%%%%%%%%%%%%%%%%%%%%%%%%%%

\section{Introduction}

The particle discovered in the Higgs-boson searches by 
ATLAS~\cite{ATLASdiscovery} and CMS~\cite{CMSdiscovery} at CERN shows, 
within experimental and theoretical uncertainties, properties compatible 
with the Higgs boson of the Standard Model 
(SM)~\cite{Aad:2015zhl,Atlas2015008,Khachatryan:2014jba}.  It can also 
be interpreted as the Higgs boson of extended models, however, where the 
lightest Higgs boson of the Minimal Supersymmetric Standard Model 
(MSSM)~\cite{mssm} is a prime candidate.

The Higgs sector of the MSSM with two scalar doublets accommodates five 
physical Higgs bosons.  In lowest order these are the light and heavy 
$\cp$-even $h$ and $H$, the $\cp$-odd $A$, and the charged Higgs bosons 
$H^\pm$.  At tree level, the Higgs sector can be parameterized 
in terms of the gauge couplings, the mass of the $\cp$-odd Higgs boson, 
$\MA$, and $\tb \equiv v_2/v_1$, the ratio of the two vacuum expectation 
values; all other masses and mixing angles follow as predictions. 

Higher-order contributions can give large corrections to the tree-level 
relations~~\cite{MHreviews,PomssmRep}, and in particular to the mass of 
the lightest Higgs boson, $\Mh$.  For the MSSM\footnote{%
	We concentrate here on the case with real parameters.  For the 
	case of complex parameters 
	see~\citeres{mhcMSSMlong,mhcMSSM2L,Demir,mhiggsCPXRG1,mhiggsFDalt2} 
	and references therein.} 
with real parameters the status of higher-order corrections to the 
masses and mixing angles in the neutral Higgs sector is quite advanced, 
see \citeres{ERZ,mhiggsf1lA,mhiggsf1lB,mhiggsf1lC} for the calculations 
of the full one-loop level.  At the two-loop 
level~\cite{mhiggsletter,mhiggslong,mhiggslle,mhiggsFDalbals,bse,mhiggsEP0,mhiggsEP1,mhiggsEP1b,mhiggsEP2,mhiggsEP3,mhiggsEP3b,mhiggsEP4,mhiggsEP4b,mhiggsRG1,mhiggsRG1a,mhiggsFDalt2} 
in particular the \order{\alt\als} and \order{\alt^2} contributions 
($\alt \equiv h_t^2 / (4 \pi)$, $h_t$ being the top-quark Yukawa 
coupling) to the self-energies -- evaluated in the Feynman-diagrammatic 
(FD) as well as in the effective potential (EP) method -- as well as the 
\order{\alb\als}, \order{\alt\alb} and \order{\alb^2} contributions -- 
evaluated in the EP approach -- are known for vanishing external 
momenta.  An evaluation of the momentum dependence at the two-loop level 
in a pure \DRbar\ calculation was presented in \citere{mhiggs2lp2}.  The 
latest status of the momentum-dependent two-loop corrections will be 
discussed below.  A (nearly) full two-loop EP calculation, including 
even the leading three-loop corrections, has also been 
published~\cite{mhiggsEP5}.  Within the EP method all contributions are 
evaluated at zero external momentum, however, in contrast to 
the FD method which in principle allows for non-vanishing external 
momenta.  Furthermore, the calculation presented in \citere{mhiggsEP5} 
is not publicly available as a computer code for Higgs-boson mass 
calculations.  Subsequently, another leading three-loop calculation of 
\order{\alt\als^2}, depending on the various SUSY mass hierarchies, was 
completed~\cite{mhiggsFD3l}, resulting in the code \texttt{H3m} which 
adds the three-loop corrections to the
\fh~\cite{feynhiggs,mhiggslong,mhiggsAEC,mhcMSSMlong,Mh-logresum} 
result.  Most recently, a combination of the full one-loop result, 
supplemented with leading and subleading two-loop corrections evaluated 
in the FD/EP method and a resummation of the leading and subleading 
logarithmic corrections from the scalar-top sector has been 
published~\cite{Mh-logresum} in the latest version of the code~\fh. 
%~\cite{feynhiggs,mhiggslong,mhiggsAEC,mhcMSSMlong,Mh-logresum}.

\medskip

The measured mass value of the observed Higgs-boson is currently known 
to about $250 \mev$ accuracy~\cite{Aad:2015zhl}, reaching the level of a 
precision observable.  At a future linear collider (ILC), the precise 
determination of the light Higgs-boson properties and/or heavier MSSM 
Higgs-bosons within the kinematic reach will be 
possible~\cite{Snowmass13HiggsWP}.  In particular, a mass measurement of 
the light Higgs-boson with an accuracy below $\sim 0.05 \gev$ is 
anticipated~\cite{dbd}.

In \citere{mhiggsAEC} the remaining theoretical uncertainty in the 
calculation of $\Mh$, from unknown higher-order corrections, was 
estimated to be up to $3 \gev$, depending on the parameter region; see 
also \citeres{Mh-logresum,ehowp} for updated results. As the accuracy of 
the $\Mh$ prediction should at least match the one of the experimental 
result, higher-order corrections which do not dominate the size of the 
Higgs-boson mass values have to be included in the Higgs-boson mass 
predictions.

To better control the size of momentum-dependent contributions, we 
recently presented the calculation of the \order{p^2\alt\als} 
corrections to $\Mh$ (the leading momentum-dependent two-loop QCD 
corrections). The calculation was performed in a hybrid on-shell/\DRbar\ 
scheme~\cite{Mh-p2-BH4} at the two-loop level, where $\MA$ and the 
tadpoles are renormalized on-shell (OS), whereas the Higgs-boson fields 
and $\tb$ are renormalized \DRbar.  At the one-loop level the top/stop 
parameters are renormalized OS.\footnote{%
	From a technical point of view we calculated the
	momentum-dependent two-loop self-energy diagrams numerically 
	using the program 
	\sd~\cite{Carter:2010hi,Borowka:2012yc,Borowka:2015mxa}.}
Subsequently, in \citere{Mh-p2-DDVS} this calculation was repeated with 
a different result (also, a calculation in a pure \DRbar\ scheme as well 
as the two-loop corrections of \order{\al\als} were presented).  Within 
\citere{Mh-p2-DDVS} the discrepancy between \citeres{Mh-p2-BH4} and 
\cite{Mh-p2-DDVS} was explained by an inconsistency in the 
renormalization scheme used for the Higgs-boson field renormalization in 
\citere{Mh-p2-BH4}.

In this paper we demonstrate that this claim is incorrect.  The 
renormalization scheme for the Higgs-boson fields used in 
\citere{Mh-p2-BH4} is (up to corrections beyond the two-loop level) 
identical to the one employed in \citere{Mh-p2-DDVS}.  We clarify that 
the differences between the two results originates in a difference of 
the top-quark-mass renormalization scheme.  While in \citere{Mh-p2-BH4} 
a full OS renormalization was used, in \citere{Mh-p2-DDVS} the 
contributions to the top-quark self-energy of \order{\eps} (with $4 - D 
= 2\eps$, $D$ being the space-time dimension) were neglected, leading to 
the observed numerical differences.
We also demonstrate how this difference in
the treatment of the contributions from the top quark mass can be linked to
a difference in the two-loop field renormalization constant
and explain why this difference should be regarded as a theoretical
uncertainty at the two-loop level,
which would be fixed only at three loop order.

We further present a consistent calculation of the \order{p^2\alt\als} 
corrections to $\Mh$ in a scheme where the top quark is renormalized 
\DRbar, whereas the scalar tops continue to be renormalized OS.  This 
new scheme is available from \fh\ version 2.11.1 on, allowing for 
an improved estimate of (some) unknown higher-order corrections 
beyond the two-loop level originating from the top/stop sector.

The paper is organized as follows.  An overview of the relevant sectors 
and the renormalization employed in our calculation is given in 
\refse{sec:renorm}.  In \refse{sec:comp} we compare analytically and 
numerically the results of \citeres{Mh-p2-BH4} and \cite{Mh-p2-DDVS}.  
Results obtained using the \DRbar\ scheme for the top-quark mass are 
given in \refse{sec:drbar}.  Our conclusions are given in 
\refse{sec:conclusions}.

%%%%%%%%%%%%%%%%%%%%%%%%%%%%%%%%%%%%%%%%%%%%%%%%%%%%%%%%%%%%%%%%%%%%%%%%%%%%%%%
%%%%%%%%%%%%%%%%%%%%%%%%%%%%%%%%%%%%%%%%%%%%%%%%%%%%%%%%%%%%%%%%%%%%%%%%%%%%%%%

\section{The relevant sectors and their renormalization}
\label{sec:renorm}

\subsection{The Higgs-boson sector of the MSSM}
\label{sec:higgs}

The MSSM requires two scalar doublets, which are conventionally written 
in terms of their components as follows,
\begin{align*}
\cHe &= \VL \cHe^0 \\[0.5ex] \cHe^- \VR = \VL v_1 
      + \frac{1}{\sqrt2}(\phi_1^0 - i\chi_1^0) \\[0.5ex] -\phi_1^- \VR\,,
        \\
\cHz &= \VL \cHz^+ \\[0.5ex] \cHz^0 \VR = \VL \phi_2^+ \\[0.5ex] 
        v_2 + \frac{1}{\sqrt2}(\phi_2^0 + i\chi_2^0) \VR\,.
\end{align*}
The bilinear part of the Higgs potential leads to the tree-level mass 
matrix for the neutral $\cp$-even Higgs-bosons,
\begin{align}
\label{higgsmassmatrixtree}
M_{\text{Higgs}}^{2,\text{tree}} = \ML \mpe^2 & \mpez^2 \\ 
                           \mpez^2 & \mpz^2 \MR  = 
      \ML \MA^2 \SQb + \MZ^2 \CQb & -(\MA^2 + \MZ^2) \Sbe \Cb \\ 
    -(\MA^2 + \MZ^2) \Sbe \Cb & \MA^2 \CQb + \MZ^2 \SQb \MR ,
\end{align}
in the $(\Pe, \Pz)$ basis, expressed in terms of the 
$Z$~boson mass, $M_Z$, 
$M_A$ and the angle $\beta$. Diagonalization via the angle $\al$ yields 
the tree-level masses $\mhtree$ and $\mHtree$.
Below we also use $\MW$, denoting the $W$~boson mass and $\sw$,
the sine of the weak mixing angle, 
$\sw = \sqrt{1 - \cw^2} = \sqrt{1 - \MW^2/\MZ^2}$.

\bigskip

The higher-order-corrected $\cp$-even Higgs-boson masses in the MSSM are 
obtained from the corresponding propagators dressed by their 
self-energies.  The calculation of these and their renormalization is 
performed in the $(\Pe, \Pz)$ basis, which has the advantage that the 
mixing angle $\al$ does not appear and expressions are in general 
simpler.  The inverse propagator matrix in the $(\Pe, \Pz)$ basis is 
given by
\begin{align}
\label{eq:prop}
(\Delta_{\text{Higgs}})^{-1} = -\text{i}
\left( \begin{matrix} 
p^2 - m_{\phi_1}^2 + \hat{\Sigma}_{\phi_1}(p^2) & -m_{\phi_1\phi_2}^2 +\hat{\Sigma}_{\phi_1\phi_2}(p^2)\\ 
-m_{\phi_1\phi_2}^2 +\hat{\Sigma}_{\phi_1\phi_2}(p^2) & p^2 - m_{\phi_2}^2 + \hat{\Sigma}_{\phi_2}(p^2) 
\end{matrix} \right),
\end{align}
where $\hat{\Sigma}(p^2)$ denote the renormalized Higgs-boson 
self-energies, $p$ being the external momentum.  The renormalized 
self-energies can be expressed through the unrenormalized self-energies, 
$\Si(p^2)$, and counterterms involving renormalization constants $\delta 
m^2$ and $\delta Z$ from parameter and field renormalization.  With the 
self-energies expanded up to two-loop order, $\hSi = \hSi^{(1)} + 
\hSi^{(2)}$, one has for the $\cp$-even part at the $i$-loop level ($i = 
1, 2$),
\begin{subequations}
\label{rMSSM:renses_higgssector}
\begin{align}
\label{rse11}
\seri{\Pe}(p^2)  &=\, \sei{\Pe}(p^2) + \dZi{\Pe}\, (p^2-\mpe^2) - \dmesqi~, \\
\label{rse12}
\seri{\PePz}(p^2)  &=\, \sei{\PePz}(p^2) - \dZi \PePz \, \mpez^2 - \dmezsqi~,\\
\label{rse22}
\seri{\Pz}(p^2)  &=\, \sei{\Pz}(p^2) + \dZi{\Pz}\, (p^2-\mpz^2) - \dmzsqi~.
\end{align}
\end{subequations}
At the two-loop level the expressions in
\refeqs{rMSSM:renses_higgssector} do not contain contributions of 
the type (1-loop)~$\times$~(1-loop); such terms do 
not appear at \order{\alt\als} and hence can be omitted in the context 
of this paper.  For the general expressions see 
\citere{mhiggsFDalt2}.

Beyond the one-loop level, unrenormalized self-energies contain sub-loop 
renormalizations.  At the two-loop level, these are one-loop diagrams 
with counterterm insertions at the one-loop level.

%%%%%%%%%%%%%%%%%%%%%%%%%%%%%%%%%%%%%%%%%%%%%%%%%%%%%%%%%%%%%%%%%%%%%%%%%%%%%%

\subsection{Renormalization}
\label{sec:renorm-det}

The following section summarizes the renormalization worked out in 
\citere{Mh-p2-BH4}, based on \citere{mhiggslong}.
%
%\medskip
%
The field renormalization is carried out by assigning one 
renormalization constant to each doublet,
\begin{align}
\label{rMSSM:HiggsDublettFeldren}
  \cHe \to (1 + \tfrac{1}{2} \dZ{\cHe} )\, \cHe, \quad
  \cHz \to (1 + \tfrac{1}{2} \dZ{\cHz} ) \cHz\,,
\end{align} 
which can be expanded to one- and two-loop order according to
\begin{align}
\label{rMSSM:Feldrenexpand}
 \dZ{\cHe} & =\, \dZo{\cHe} +  \dZt{\cHe} \, , \quad
  \dZ{\cHz}  \, =\,  \dZo{\cHz} + \dZt{\cHz}\,. 
\end{align}
The field renormalization constants appearing in~(\ref{rMSSM:renses_higgssector})
are then given by
\begin{align}
\label{fieldrenconstphi}
\dZ{\Pe}^{(i)} = \dZ{\cHe} ^{(i)}\,, \quad 
\dZ{\Pz}^{(i)} = \dZ{\cHz} ^{(i)}\, , \quad
\dZ{\Pe\Pz}^{(i)} = \edz (\dZ{\cHe}^{(i)} + \dZ{\cHz}^{(i)}  )   \, .
\end{align}
The mass counterterms $\delta m^{2 (i)}_{ab}$ in 
\refeq{rMSSM:renses_higgssector} are derived from the Higgs 
potential, including the tadpoles, by the following parameter 
renormalization,
\begin{align}
\label{rMSSM:PhysParamRenorm}
  \MA^2 &\to \MA^2 + \dMAsqo + \dMAsqt,  
& \tade &\to \tade + \dtadeo + \dtadet, \\ 
  \MZ^2 &\to \MZ^2 + \dMZsqo + \dMZsqt,  
& \tadz &\to \tadz + \dtadzo + \dtadzt, \notag \\ 
     \tb & \to \tb \KL 1 + \de\tb^{(1)} + \de\tb^{(2)} \KR~. \notag
\end{align}
The parameters $\tade$ and $\tadz$ are the terms linear in $\Pe$ and 
$\Pz$ in the Higgs potential.  The renormalization of the $Z$-mass $M_Z$ 
does not contribute to the $\mathcal{O}(\alpha_s \alpha_t)$ corrections 
we are pursuing here; it is listed for completeness only. 

\medskip

The basic renormalization constants for parameters and fields have to be 
fixed by renormalization conditions according to a renormalization 
scheme.  Here we choose the on-shell scheme for the parameters and the 
\drbar\ scheme for field renormalization and give the expressions for 
the two-loop part.  This is consistent with the renormalization scheme 
used at the one-loop level.

\medskip

The tadpole coefficients are chosen to vanish at all orders; hence
their two-loop counterterms follow from  
\begin{align}
T_{1,2}^{(2)}+ \de T_{1,2}^{(2)} = 0\,, \quad\text{\ie}\quad
  \dtadet = -{\tadet}, \quad \dtadzt = -{\tadzt}\,,
\end{align}
where $\tadet$, $\tadzt$ are obtained from the two-loop tadpole 
diagrams.  The two-loop renormalization constant of the $A$-boson mass 
reads
\begin{align}
\label{rMSSM:mass_osdefinition}
\dMAsqt = \re \se{AA}^{(2)}(\MA^2) ,
\end{align}
in terms of the $A$-boson unrenormalized self-energy $\se{AA}$.  The 
appearance of a non-zero momentum in the self-energy goes beyond the 
\order{\alt\als} corrections evaluated in 
\citeres{mhiggsletter,mhiggslong,mhiggsEP1}.

\medskip

For the renormalization constants $\dZ{\cHe}$, $\dZ{\cHz}$ and $\de\tb$ 
several choices are possible, see the discussion in~\cite{tbren}.  As 
shown there, the most convenient choice is a \drbar\ renormalization of 
$\de\tb$, $\dZ{\cHe}$ and $\dZ{\cHz}$, which at the two-loop level reads
\begin{subequations}
\label{rMSSM:deltaZHiggsTB}
\begin{align}
  \dZ{\cHe}^{(2)} &= \dZ{\cHe}^{\DRbar (2)}
       \; = \; - \KKL \re \Sipt_{\Pe} \KKR^{\text{div}}_{|p^2 = 0}\,, \\[.5em]
  \dZ{\cHz}^{(2)} &= \dZ{\cHz}^{\DRbar (2)} 
       \; = \; - \KKL \re \Sipt_{\Pz} \KKR^{\text{div}}_{|p^2 = 0}\,, \\[.5em]
  \dtanbt &= \dtanb^{\DRbar (2)} 
       \; = \; \edz  \KL \dZ{\cHz}^{(2)} - \dZ{\cHe}^{(2)} \KR\,.
       \label{eq:tanbren}
\end{align}
\end{subequations}
The term in \refeq{eq:tanbren} is in general not the proper expression 
beyond one-loop order even in the \drbar\ scheme. For our approximation, 
however, with only the top Yukawa coupling at the two-loop level, it is 
the correct \drbar\ form~\cite{Sperling:2013eva}.

\medskip

The two-loop mass counterterms in the renormalized 
self-energies~(\ref{rMSSM:renses_higgssector}) are now expressed in 
terms of the two-loop parameter renormalization constants, determined 
above, as follows,
\begin{subequations}
\label{masscounterterms}
\begin{align}
%V11:
\label{dm1sq}
\dmesqt &= \,\dMZsqt \, \CQb + \dMAsqt \SQb \\
\nonumber &\quad - \dtadet \frac{e}{2 \MW \sw} \, \Cb (1 + \SQb) 
+ \dtadzt \frac{e}{2 \MW \sw} \, \CQb \Sbe \\
\nonumber &\quad 
+ 2\, \dtanbt \, \cos^2\!\beta \sin^2\!\beta \, (\MA^2 - \MZ^2)\,, 
\\[1ex]
%V12:
\label{dm12sq}
\dmezsqt &= - (\dMZsqt + \dMAsqt) \Sbe \Cb \\
\nonumber &\quad - \dtadet \frac{e}{2 \MW \sw} \, \SDb 
                                - \dtadzt \frac{e}{2 \MW \sw} \, \CDb \\
\nonumber &\quad 
                 -  \dtanbt\, \cos\beta \sin\beta \cos 2\beta \, 
                      (\MA^2 + \MZ^2)\,, \\[1ex]
%V22:
\label{dm2sq}
\dmzsqt &= \dMZsqt \SQb + \dMAsqt \CQb \\
\nonumber &\quad + \dtadet \frac{e}{2 \MW \sw} \, \SQb \cos\beta
                 - \dtadzt \frac{e}{2 \MW \sw} \, \Sbe (1 + \CQb) \\
\nonumber &\quad - 2\, \dtanbt \, \cos^2\!\beta \sin^2\!\beta \, (\MA^2 - \MZ^2)\,.
\end{align}
\end{subequations}
The $Z$-mass counterterm is again kept for completeness; it does not 
contribute in the approximation of \order{\alt\als} considered here.

%%%%%%%%%%%%%%%%%%%%%%%%%%%%%%%%%%%%%%%%%%%%%%%%%%%%%%%%%%%%%%%%%%%%%%%%%%%%%

\subsection{Diagram evaluation}

\begin{figure}[htb!]
%\vspace{-1em}
\begin{center}
\hspace{-120pt}
\subfigure[]{\raisebox{0pt}{\includegraphics[width=0.2\textwidth]{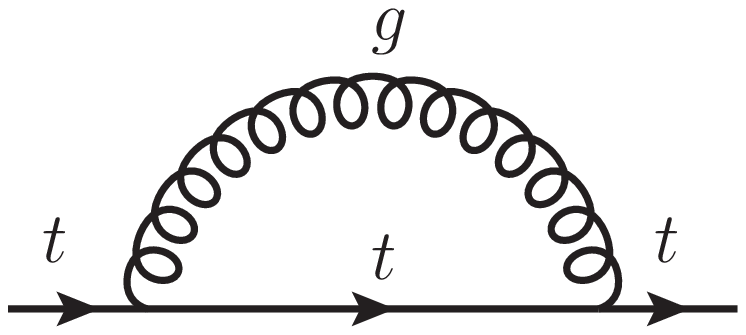}} \label{subfig:b}}
\hspace{15pt}
\subfigure[]{\raisebox{0pt}{\includegraphics[width=0.2\textwidth]{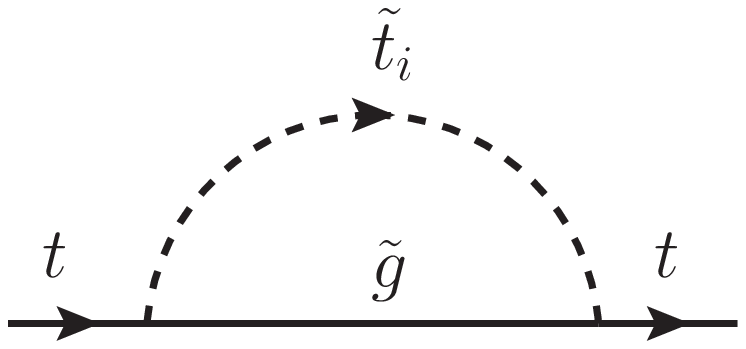}} \label{subfig:d}}\\
\subfigure[]{\raisebox{0pt}{\includegraphics[width=0.2\textwidth]{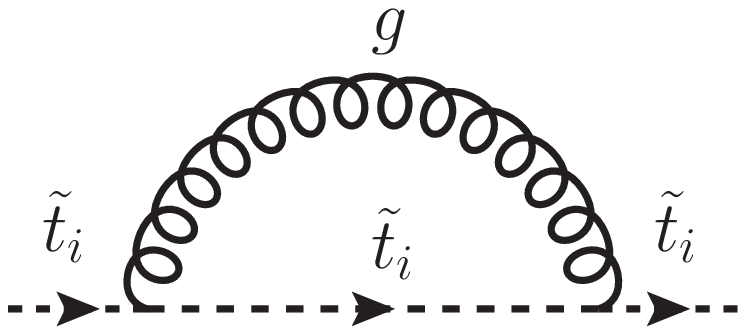}} \label{subfig:a}}
\hspace{15pt}
\subfigure[]{\raisebox{0pt}{\includegraphics[width=0.2\textwidth]{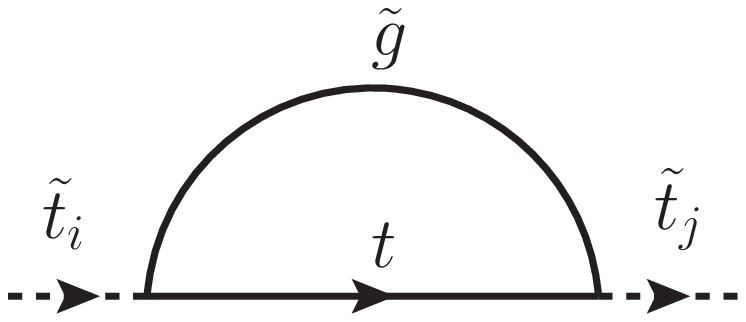}} \label{subfig:c}}
\hspace{15pt}
\subfigure[]{\raisebox{1pt}{\includegraphics[width=0.2\textwidth]{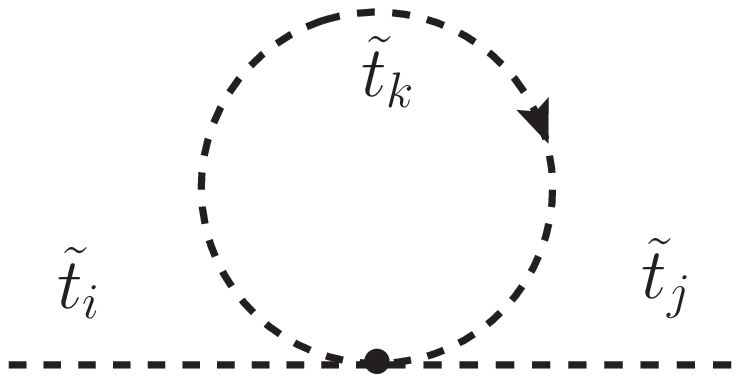}} \label{subfig:e}}
\caption{Generic one-loop diagrams for subrenormalization counterterms
for the top quark (upper row) and for the scalar tops (lower row)
%, involving top quarks $t$,
%top squarks~$\tilde{t}$, gluons $g$ and gluinos $\tilde{g}$  
($i,j,k =1,2$). }
\label{fig:fd_ctis}
\end{center}
%\vspace{-3em}
\end{figure}

Our calculation is performed in the Feynman-diagrammatic (FD) approach.  
To arrive at expressions for the unrenormalized self-energies and 
tadpoles at \order{\alt\als}, the evaluation of genuine two-loop 
diagrams and one-loop graphs with counterterm insertions is required. 
For the counterterm insertions, described in subsection~\ref{sec:stop}, 
one-loop diagrams with external top quarks/squarks have to be evaluated 
as well, as displayed in~\reffi{fig:fd_ctis}.  The calculation is 
performed in dimensional reduction~\cite{dred}.

The complete set of contributing Feynman diagrams was generated 
with the program \texttt{FeynArts}~\cite{feynarts} (using the model file 
including counterterms from \citere{mssmct}), tensor reduction and the 
evaluation of traces was done with support from the programs 
\texttt{FormCalc}~\cite{formcalc} and \texttt{TwoCalc}~\cite{twocalc}, 
yielding algebraic expressions in terms of the scalar one-loop functions 
$A_0$, $B_0$~\cite{oneloop}, the massive vacuum two-loop 
functions~\cite{Davydychev:1992mt}, and two-loop integrals which depend 
on the external momentum.  These integrals were evaluated with the 
program \sd~\cite{Carter:2010hi,Borowka:2012yc,Borowka:2015mxa}, where 
up to four different masses in 34 different mass configurations needed 
to be considered, with differences in the kinematic invariants of 
several orders of magnitude.

%%%%%%%%%%%%%%%%%%%%%%%%% F I G U R E %%%%%%%%%%%%%%%%%%%%%%%%%%%%%%%%%%%%%%%%%
%%%%%%%%%%%%%%%%%%%%%%%%% F I G U R E %%%%%%%%%%%%%%%%%%%%%%%%%%%%%%%%%%%%%%%%%

%%%%%%%%%%%%%%%%%%%%%%%%%%%%%%%%%%%%%%%%%%%%%%%%%%%%%%%%%%%%%%%%%%%%%%%%%%%%%%%

\subsection{The scalar-top sector of the MSSM}
\label{sec:stop}

The bilinear part of the top-squark Lagrangian,
\begin{align}
\cL_{\Stop, \text{mass}} &= - \begin{pmatrix}
{{\tilde{t}}_{L}}^{\dagger}, {{\tilde{t}}_{R}}^{\dagger} \end{pmatrix}
\matr{M}_{\tilde{t}}\begin{pmatrix}{\tilde{t}}_{L}\\{\tilde{t}}_{R}
\end{pmatrix} \,,
\end{align}
contains the stop-mass matrix
\begin{align}\label{Sfermionmassenmatrix}
\matr{M}_{\tilde{t}} 
&= \begin{pmatrix}  
 \MstL^2 + \mt^2 + \MZ^2 \CZb \, (T_t^3 - Q_t \sw^2) & 
 \mt \Xt \\[.2em]
 \mt \Xt &
 \MstR^2 + \mt^2 + \MZ^2 \CZb \, Q_t \, \sw^2
\end{pmatrix},
\intertext{with}
\Xt &= \At - \mu\,\CTb
\end{align}
where $Q_t$ and $T_t^3$ denote the charge and isospin of the top quark, 
$\At$ the trilinear coupling between the Higgs bosons and the scalar 
tops, and $\mu$ the Higgsino mass parameter.  Below we use $\msusy := 
\MstL = \MstR$ for our numerical evaluation. The analytical calculation 
was performed for arbitrary $\MstL$ and $\MstR$, however. 
$\matr{M}_{\tilde{t}}$ can be diagonalized with the help of a unitary 
transformation matrix ${\matr{U}}_{\tilde t}$, parameterized by a mixing 
angle ${\theta}_{\tilde{t}}$, to provide the eigenvalues $\mste^2$ and 
$\mstz^2$ as the squares of the two on-shell top-squark masses.

\medskip

For the evaluation of the \order{\alt\als} two-loop contributions to the 
self-energies and tadpoles of the Higgs sector, renormalization of the 
top/stop sector at \order{\als} is required, giving rise to the 
counterterms for sub-loop renormalization.
%(see \reffis{fig:fd_hHA},\ref{fig:fd_TP}). 
We follow the renormalization at the one-loop level given in 
\citeres{mhiggsFDalbals,hr,SbotRen,Stop2decay}, where details can be 
found.  In particular, in the context of this paper, an OS 
renormalization is performed for the top-quark mass as well as for the 
scalar-top masses.  This is different from the approach pursued, for 
example, in~\citere{mhiggs2lp2}, where a \DRbar\ renormalization was 
employed, or similarly in the pure \DRbar\ renormalization presented in 
\citere{Mh-p2-DDVS}.  Using the OS scheme allows us to consistently 
combine our new correction terms with the hitherto available 
self-energies included in \fh.

Besides employing a pure OS renormalization for the top/stop masses in 
our calculation, we also obtain a result in which the top-quark mass is 
renormalized \DRbar.  This new top-quark mass renormalization is 
included as a new option in the code \fh.  The comparison of the results 
using the \DRbar\ and the OS renormalization allows to estimate (some) 
missing three-loop corrections in the top/stop sector.

\medskip

Finally, at \order{\alt\als}, gluinos appear as virtual particles only 
at the two-loop level (hence, no renormalization for the gluinos is 
needed).  The corresponding soft-breaking gluino mass parameter $M_3$ 
determines the gluino mass, $\mgl = M_3$.

%%%%%%%%%%%%%%%%%%%%%%%%%%%%%%%%%%%%%%%%%%%%%%%%%%%%%%%%%%%%%%%%%%%%%%%%%%

\subsection{Evaluation and implementation in the program \fh}
\label{sec:fh}

The resulting new contributions to the neutral $\cp$-even Higgs-boson 
self-energies, containing all momentum-dependent and additional constant 
terms, are assigned to the differences
\begin{equation}
\label{eq:DeltaSE}
\De\ser{ab}(p^2) = \ser{ab}^{(2)}(p^2) - \tilde\Sigma_{ab}^{(2)}(0)\,,
\qquad
ab = \{HH,hH,hh\}\,.
\end{equation}
These are the new terms evaluated in \citere{Mh-p2-BH4}, included in 
\fh.  Note the tilde (not hat) on $\tilde\Sigma^{(2)}(0)$ which 
signifies that not only the self-energies are evaluated at zero external 
momentum but also the corresponding counterterms, following 
\citeres{mhiggsletter,mhiggslong}.  A finite shift $\De\hat{\Sigma}(0)$ 
therefore remains in the limit $p^2\to 0$ due to $\dMAsqt = 
\re\se{AA}^{(2)}(\MA^2)$ being computed at $p^2=\MA^2$ in 
$\hat\Sigma^{(2)}$, but at $p^2=0$ in $\tilde\Sigma^{(2)}$; for details 
see \refeqs{rMSSM:mass_osdefinition} and (\ref{masscounterterms}).  For 
the sake of simplicity we will refer to these terms as \order{p^2 \alt 
\als} despite the $\MA^2$ dependence.

%%%%%%%%%%%%%%%%%%%%%%%%%%%%%%%%%%%%%%%%%%%%%%%%%%%%%%%%%%%%%%%%%%%%%%%%%%
%%%%%%%%%%%%%%%%%%%%%%%%%%%%%%%%%%%%%%%%%%%%%%%%%%%%%%%%%%%%%%%%%%%%%%%%%%

\section{Discussion of renormalization schemes}
\label{sec:comp}

In this section we compare our results for the \order{p^2 \alt \als} 
contributions to the MSSM Higgs-boson self-energies, as given in 
\citere{Mh-p2-BH4} to the ones presented subsequently in 
\citere{Mh-p2-DDVS}.  We first show analytically the agreement in the 
Higgs field renormalization in the two calculations and discuss the 
differences in the $\mt$ renormalizations.  We also present some 
numerical results in both schemes, demonstrating agreement with 
\citere{Mh-p2-DDVS} once the \order{\eps} terms are dropped from the 
top-quark mass counterterm.

Using an OS renormalization for the top-quark mass, the counterterm is 
determined from the components of the \order{\alpha_s} top-quark 
self-energy (Fig.~\ref{fig:fd_ctis}) as follows,
\begin{align}
\label{dmtOS}
\frac{\dmtOS}{\mt} &= \ed{2} \re \KKKL 
     \KKL\Si_t^L (\mt^2) + \Si_t^R (\mt^2) \KKR  
  + \KKL \Si_t^{SL} (\mt^2) + \Si_t^{SR} (\mt^2) \KKR \KKKR,
\end{align}
where the top-quark self-energy is decomposed according to
\begin{align}
\label{decomposition}
\Si_t (p) &= \not\! p\, \omega_{-} \Si_t^L (p^2)
                   + \not\! p\, \omega_{+} \Si_t^R (p^2)
                   + \mt\,\omega_{-} \Si_t^{SL} (p^2) 
                   + \mt\,\omega_{+} \Si_t^{SR} (p^2)~.
\end{align}
with the projectors $\omega_{\pm} = \frac{1}{2}(\unity\pm\gamma_5)$.

%%%%%%%%%%%%%%%%%%%%%%%%%%%%%%%%%%%%%%%%%%%%%%%%%%%%%%%%%%%%%%%%%%%%%%%%%

\subsection{Analytical comparison}

\begin{figure}[t]
\begin{center}
\includegraphics[width=0.2\textwidth]{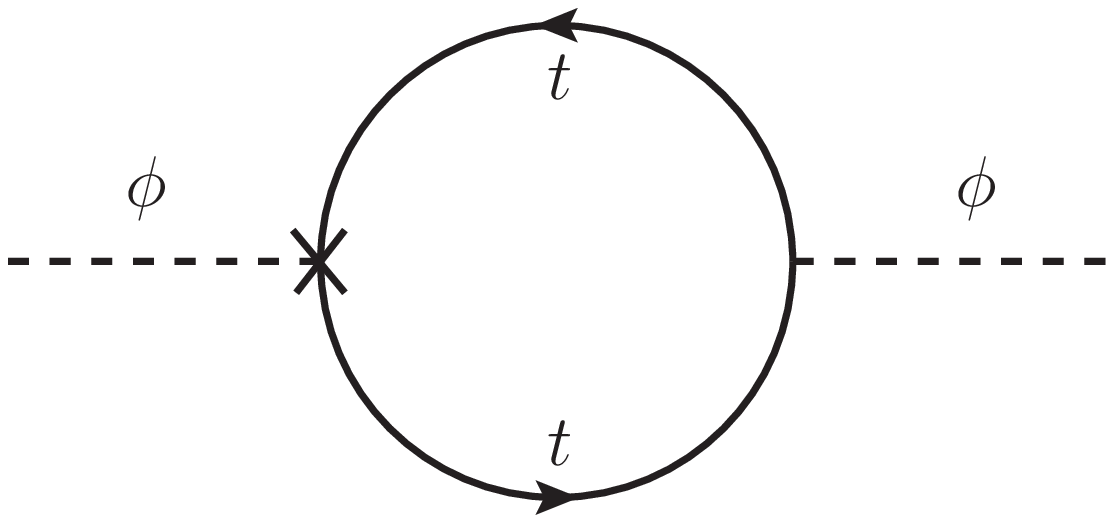}\qquad
\includegraphics[width=0.2\textwidth]{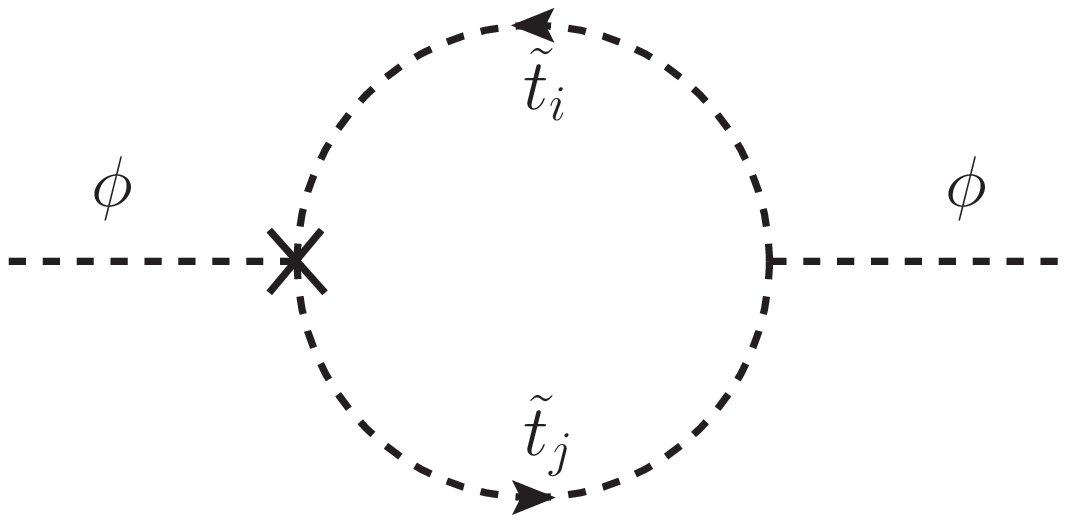}
\caption{One-loop subrenormalization diagram contributing to
  $\de_{\Si_{22}}(p^2)$ and $\de_A(p^2)$, with the counterterm insertion
  denoted by a cross.  The right diagram only contributes to 
  $\de_{\Si_{22}}(0)$ and $\de_A(0)$.}
\label{fig:relevant1Lct}
\end{center}
\end{figure}

In the \order{\alt\als} calculation of the Higgs-boson self-energies the 
renormalization of the top-quark mass at \order{\als} is required.  The 
contributing diagrams are shown in the top row of \reffi{fig:fd_ctis}.  
The top-quark mass counterterm is inserted into the sub-loop 
renormalization of the two-loop contributions to the Higgs-boson 
self-energies, where two sample diagrams are shown in 
\reffi{fig:relevant1Lct}.  The left diagram contributes to the 
momentum-dependent two-loop self-energies, while the right one 
contributes only to the momentum-independent part.
Evaluating the expression in \refeq{dmtOS} in $4 - 2\eps$ dimensions yields 
the OS top-quark mass counterterm at the one-loop level, which can be 
written as a Laurent expansion in $\eps$,
\begin{align}
\dmtOS &= \ed{\eps}\,\dmtdiv + \dmtfin + \eps\,\dmteps + \ldots\,;
\label{dmt}
\end{align}
higher powers in $\varepsilon$, indicated by the ellipses, do not 
contribute at the two-loop level for $\varepsilon\to 0$ after 
renormalization.  Accordingly, the \DRbar\ top-quark mass counterterm is 
given by the singular part of \refeq{dmt},
\begin{align}
\dmtDRbar &= \ed{\eps}\,\dmtdiv\,.
\label{dmtdrbar}
\end{align}

%%%%%%%%%%%%%%%%%%%%%%%%% F I G U R E %%%%%%%%%%%%%%%%%%%%%%%%%%%%%%%%%%%%%%%%%
%\begin{figure}[ht!]
%%%%%%%%%%%%%%%%%%%%%%%%% F I G U R E %%%%%%%%%%%%%%%%%%%%%%%%%%%%%%%%%%%%%%%%%

\noindent
For further use we define the quantity 
\begin{align}
\dmtFIN &= \ed{\eps}\,\dmtdiv + \dmtfin\,.
\label{dmtfin}
\end{align}
At \order{\als} the OS counterterm is given as
\begin{align}
\frac{\dmtOS}{\mt} &=\frac{\als}{6 \pi} \, \Bigg\{
- 2\frac{A_0(\mt^2)}{\mt^2}-4\,B_0(\mt^2,0,\mt^2) \non \\
&\quad -2 \frac{A_0(\mgl^2)}{\mt^2} + \frac{A_0(\mste^2)}{\mt^2}
+ \frac{A_0(\mstz^2)}{\mt^2} \non\\
&\quad +
\frac{\mgl^2 +\mt^2 -\mste^2  - 4 \,\sin\tst \cos\tst\,\mgl\, \mt  }{\mt^2}
\,\re[B_0(\mt^2,\mgl^2,\mste^2)] \non\\
&\quad +
\frac{\mgl^2 +\mt^2 -\mstz^2  + 4 \,\sin\tst\cos\tst \,\mgl\, \mt  }{\mt^2}
\,\re[B_0(\mt^2,\mgl^2,\mstz^2)] \Bigg\} \;.
\label{dmtAB}
\end{align}
The one- and two-point functions $A_0(m^2)$ and $B_0(p^2,m_1^2,m_2^2)$ 
are expanded in $\eps$ as follows, 
\begin{align}
A_0(m^2) &= \ed{\eps}\,A_0^{\text{div}}(m^2) + A_0^{\text{fin}}(m^2) + 
            \eps\,A_0^\eps(m^2)~, \non \\
B_0(p^2,m_1^2,m_2^2) &= \ed{\eps}\,B_0^{\text{div}}(p^2,m_1^2,m_2^2) 
           + B_0^{\text{fin}}(p^2,m_1^2,m_2^2) + \eps\,B_0^\eps(p^2,m_1^2,m_2^2)~.
\end{align} 
Consequently, the term at \order{\eps}, $\dmteps/\mt$, is given by 
\refeq{dmtAB}, but taking only into account the pieces $\propto 
A_0^\eps,\, B_0^{\eps}$.  The special cases of $A_0^\eps(m^2)$ and 
$B_0^\eps(m^2,0,m^2)$ are given by
\begin{align}
A_0^\eps(m^2) &= m^2\left\{1-\log(m^2/\mu^2)+\frac{1}{2}\log^2(m^2/\mu^2)
                          + \frac{\pi^2}{12}\right\},\non\\
B_0^\eps(m^2,0,m^2) &= 4 -2\,\log(m^2/\mu^2)+\frac{1}{2}\log^2(m^2/\mu^2)
                          + \frac{\pi^2}{12}\,,
\end{align}
where the factor $4\pi\mathrm{e}^{-\ga_E}$ is absorbed into the 
renormalization scale.  The expression for $B_0^{\eps}$ depending on 
three mass scales can be found \eg in \citere{Nierste:1992wg}.

\medskip

In our calculation in \citere{Mh-p2-BH4} we include terms up to 
\order{\eps}, originating from the top-quark self-energy, in the 
top-mass counterterm\footnote{
	Taking \order{\eps} terms into account in the expressions for 
	on-shell counterterms beyond one loop is widely used in the 
	literature, see e.g.\ 
	\citeres{Melnikov:2000zc,Bonciani:2010mn,SMthresholdcorrexample}.}, 
\ie
\begin{align}
\dmt{\mbox{\tiny \cite{Mh-p2-BH4}}} = \dmtOS~.
\label{dmtBH4}
\end{align}

The derivation in \citere{Mh-p2-DDVS} proceeds differently.  The 
renormalized Higgs-boson self-energies are first calculated in a pure 
\DRbar\ scheme.  This concerns the top mass, the scalar-top masses, the 
Higgs field renormalization, and $\tb$.  In this way it is ensured that 
in particular the Higgs fields are renormalized using \DRbar, $\dZ{\cHi} 
= \dZ{\cHi}^{\DRbar}$, where this quantity contains the contribution 
from the one- and two-loop level.  Using this pure \DRbar\ scheme a 
finite result is obtained in which all poles in $1/\eps$ and $1/\eps^2$ 
cancel, such that the limit $\eps \to 0$ can be taken.  Subsequently, 
the \DRbar\ top-quark mass counterterm, $\dmtDRbar$, is replaced by an 
on-shell counterterm, and the top-quark mass definition is changed 
accordingly.  (The same procedure is applied for the scalar-top masses.)  
Since these finite expressions for the renormalized Higgs-boson 
self-energies do not contain any term of \order{1/\eps}, the $\dmteps$ 
part of the OS top-quark mass counterterm does not contribute, \ie
\begin{align}
\dmt{\mbox{\tiny \cite{Mh-p2-DDVS}}} &= \dmtFIN~.
\label{dmtDDVS}
\end{align}
The numerical results for the renormalized Higgs-boson self-energies
obtained this way differ significantly from the ones obtained in
\citere{Mh-p2-BH4}, as pointed out in \citere{Mh-p2-DDVS}.

\medskip

In the following we discuss the different Higgs-boson field 
renormalizations, where we use the notation of 
$\dZ{\cHz}^{\de\mt^{\text{X}}}$ for the field renormalization derived 
using $\de\mt^{\text{X}}$, with $\text{X} = \DRbar$, FIN, $\OS$.  
The field renormalization can be decomposed into one-loop, two-loop, 
\dots parts as
\begin{align}
\dZ{\cHz}^{\de\mt^{\text{X}}} &=
  \dZ{\cHz}^{\de\mt^{\text{X}} (1)} +
  \dZ{\cHz}^{\de\mt^{\text{X}} (2)} + \ldots
\end{align}

In \citere{Mh-p2-DDVS} it was claimed that using an OS top-quark mass
renormalization from the start results in a non-\DRbar\ renormalization
of $\dZ{\cHz}$.  While it is correct that an OS value for $\mt$ yields
different results in the one- and two-loop part, 
\begin{align}
\dZ{\cHz}^{\dmtOS (1)} \neq \dZ{\cHz}^{\dmtDRbar (1)}, \quad
\dZ{\cHz}^{\dmtOS (2)} \neq \dZ{\cHz}^{\dmtDRbar (2)}, 
\end{align}
the sum of the one- and two-loop parts are identical, independently of 
the choice of the top-quark mass renormalization (see \eg 
Eqs.~(3.60)--(3.62) in \citere{Mh-altals-NMSSM}),
\begin{align}
\KL\dZ{\cHz}^{\mbox{\tiny \cite{Mh-p2-BH4}}} = \KR\;
\dZ{\cHz}^{\dmtOS}\Big|_{\text{div}} &=
\dZ{\cHz}^{\dmtFIN} =
\dZ{\cHz}^{\dmtDRbar}
\;\KL = \dZ{\cHz}^{\mbox{\tiny \cite{Mh-p2-DDVS}}} \KR\,,
\end{align}
provided that also in $\dZ{\cHz}^{\dmtOS}$ all finite pieces are 
dropped, as done in \citere{Mh-p2-BH4}.  Differences between 
$\dZ{\cHz}^{\mbox{\tiny \cite{Mh-p2-BH4}}}$ and $\dZ{\cHz}^{\mbox{\tiny 
\cite{Mh-p2-DDVS}}}$ arise only at the three-loop level.  Consequently, 
the claim in \citere{Mh-p2-DDVS} that using $\dmtOS$ leads to an 
inconsistency in the Higgs field renormalization in \citere{Mh-p2-BH4} 
is not correct.  The field renormalizations thus cannot be responsible 
for the observed differences between \citeres{Mh-p2-BH4} and 
\cite{Mh-p2-DDVS}.

More explicitly, the difference between the two calculations results 
from non-vanishing $\dmteps$ terms in the renormalized Higgs-boson 
self-energies.  Those terms naturally appear when performing a full 
expansion in the dimensional regulator $\eps$.  The latter corresponds 
to choosing $\dmtOS$ (as done in \citere{Mh-p2-BH4}) instead of 
$\dmtFIN$ (as done in \citere{Mh-p2-DDVS}).

\medskip

In order to isolate the contributions coming from $\order{\eps}$ 
terms $\times$ $1/\eps$ poles we define the following quantities, where 
superscripts $\OS$, $\FIN$ refer to the respective use of $\dmtOS$, 
$\dmtFIN$:
\begin{subequations}
\begin{align}
\de T_i^{(2)\OS} &= \de T_i^{(2)\,\FIN} + \de_{T_i}~, \\[1ex]
\se{\phi_{ij}}^{(2)\OS}(p^2) &= \se{\phi_{ij}}^{(2)\,\FIN}(p^2) 
                            + \de_{\Sigma_{ij}}(p^2)~, \\[1ex]
\se{AA}^{(2)\OS}(p^2) &= \se{AA} ^{(2)\,\FIN} (p^2) + \de_A(p^2)~, 
\end{align}
\label{eq:deltadef}
\end{subequations}
where the last equation yields a shift for the $A$-boson mass 
counterterm in \refeq{rMSSM:PhysParamRenorm},
\begin{align}
\de\MA^{2(2)\OS} &= \de\MA^{2(2)\,\FIN}+\de_A(M_A^2)\,.
\end{align}

The $\delta$-terms are defined as the \emph{finite} contributions 
stemming from $\dmteps$-dependent parts in the counterterms (see the 
left diagram in \reffi{fig:relevant1Lct} for an example).  The 
\DRbar-renormalized quantities do not contain a finite 
$\dmteps$-dependent part by definition.  Furthermore, since $\phi_1$ has 
no coupling to the top quark, there are no terms proportional to 
$\dmteps$ in $\set{\Pe}$, $\set{\PePz}$, and $\dtadet$, and it is 
sufficient to consider $\de_{\Sigma_{22}}$, $\de_{A}$, and $\de_{T_2}$ 
only.  While $\de_{T_2}$ is $p^2$-independent, we find
\begin{align}
\de_{\Si_{22}}(p^2) &= \frac{3 \alt}{2 \pi} \,p^2 
    \frac{\dmteps}{\mt} +\de_{\Si_{22}}(0)\,,
\label{deltaterms} \\
\de_{A}(p^2) &=  \frac{3 \alt}{2 \pi} \,p^2 \CQb\,  
    \frac{\dmteps}{\mt} +\de_{A}(0)\,.
\end{align}
%Here it can be read off that $\dZ{\cHz}^{\dmtOS}$, defined as 
%$\partial\se{\phi_{2}}(p^2)/\partial p^2$, contains exactly the extra 
%term as given in \refeq{dZ2dmtOS}.  
Using 
\refeqs{rMSSM:renses_higgssector}, (\ref{masscounterterms}) we find that 
the following relations hold for the renormalized Higgs-boson 
self-energies:
\begin{align}
- \SQb\,\de_A(0) - \frac{e}{2 \MW \sw} \, \CQb\Sbe\, \de_{T_2}
&\;=\; 0~\quad (\text{for }\hSi_{\phi_1}^{(2)})\,, \non \\
\Sbe\Cb\, \de_A(0) + \frac{e}{2 \MW \sw} \, \CDb\, \de_{T_2}
&\;=\; 0~\quad (\text{for }\hSi_{\phi_1\phi_2}^{(2)})\,, \non \\
\de_{\Si_{22}}(0) - \CQb \,\de_A(0) + \frac{e}{2 \MW \sw} \, \Sbe (1 + \CQb)\, 
\de_{T_2} &\;=\; 0~\quad (\text{for }\hSi_{\phi_2}^{(2)})\,.
\label{eq:cancellation}
\end{align}
This is in agreement with the observation that in the renormalized 
Higgs-boson self-energies at zero external momentum at \order{\alt\als}, 
the terms containing $\dmteps$ drop out in the final (finite) result. 
Such a cancellation is to be expected as the same combination of 
one-loop self-energies that potentially contributes to this finite 
contribution also appears in the \order{1/\eps} term, where they must 
cancel.  This argument in principle still holds when the 
momentum-dependent \order{\alt\als} corrections are calculated and 
\emph{all} counterterms are evaluated with a full expansion in $\eps$.  
Since the counterterm $\delta_A$ is evaluated at $p^2=\MA^2$, and the 
Higgs-boson fields are renormalized in the \DRbar\ scheme, however, one 
finds, using \refeqs{rMSSM:renses_higgssector}, (\ref{masscounterterms}) 
for the three renormalized Higgs-boson self-energies,
\begin{align}
    - \SQb\, \KL \de_A(\MA^2) - \de_A(0) \KR 
&= \frac{3 \alt}{2 \pi} \KL - \cos^2\be\,\sin^2\be\,\MA^2 \KR 
\frac{\dmteps}{\mt}~\; (\mbox{for~}\hSi_{\phi_1}^{(2)})\,, \non \\
    + \Sbe\Cb\, \KL \de_A(\MA^2) - \de_A(0) \KR 
&= \frac{3 \alt}{2 \pi} \KL + \cos^3\be\,\sin\be\,\MA^2 \KR 
\frac{\dmteps}{\mt}~\; (\mbox{for~}\hSi_{\phi_1\phi_2}^{(2)})\,, \non \\
\KL \de_{\Si_{22}}(p^2) - \de_{\Si_{22}}(0) \KR 
    - \CQb\, \KL \de_A(\MA^2) - \de_A(0) \KR 
&= \frac{3 \alt}{2 \pi} \KL p^2 - \cos^4\be\,\MA^2 \KR 
\frac{\dmteps}{\mt}
~\; (\mbox{for~}\hSi_{\phi_2}^{(2)})\,, 
\label{eq:nocancellation}
\end{align}
\ie the $\dmteps$ terms contribute in the newly evaluated \order{p^2 
\alt \als} corrections.  They are $p^2$-independent in $\sert{\Pe}$ and 
$\sert{\Pe\Pz}$, while they do depend on $p^2$ in $\sert{\Pz}$.

The $p^2$-dependent terms coming from the expansion of terms like 
$(-p^2)^{-\eps}$ multiplying a $1/\eps^2$ divergence must certainly 
cancel after inclusion of the counterterms, because non-local terms 
cannot appear in a renormalizable theory. However, the cancellation of 
the $\eps$-dependent terms stemming from the mass renormalization is not 
necessarily fulfilled once the two-loop amplitude carries full momentum 
dependence.  Similarly, the truncation of the field renormalization to 
the divergent part cuts away terms involving $\dmteps$, leading to 
further non-cancellations.  The explicit \DRbar\ renormalization of the 
Higgs-boson fields drops the corresponding finite contributions, such 
that no $\dmtfin$, $\dmteps$ terms are taken into account.  The 
different dependence on the external momentum and the \DRbar\ 
prescription for the Higgs field renormalization leads to 
\refeqs{eq:nocancellation}.

\medskip 

\begin{figure}[t]
\begin{center}
\raisebox{0pt}{\includegraphics[width=0.15\textwidth]{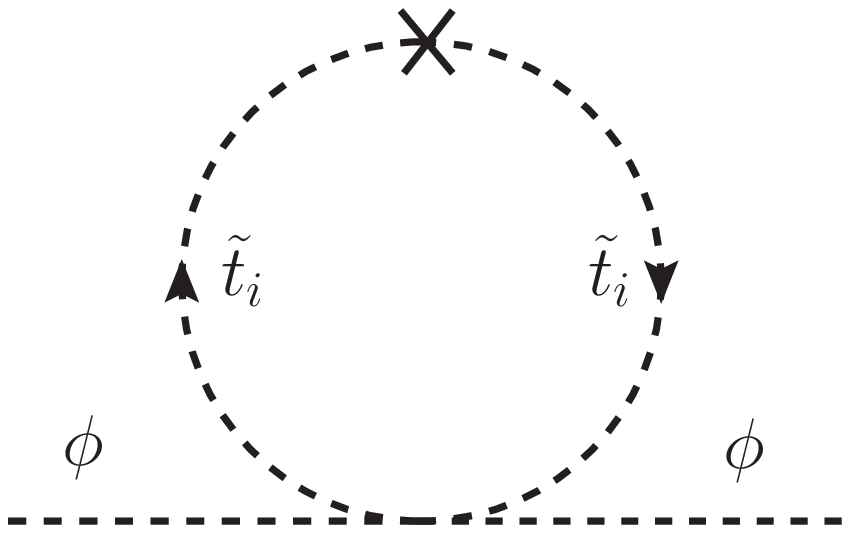}} \hspace{20pt} 
\raisebox{-2pt}{\includegraphics[width=0.2\textwidth]{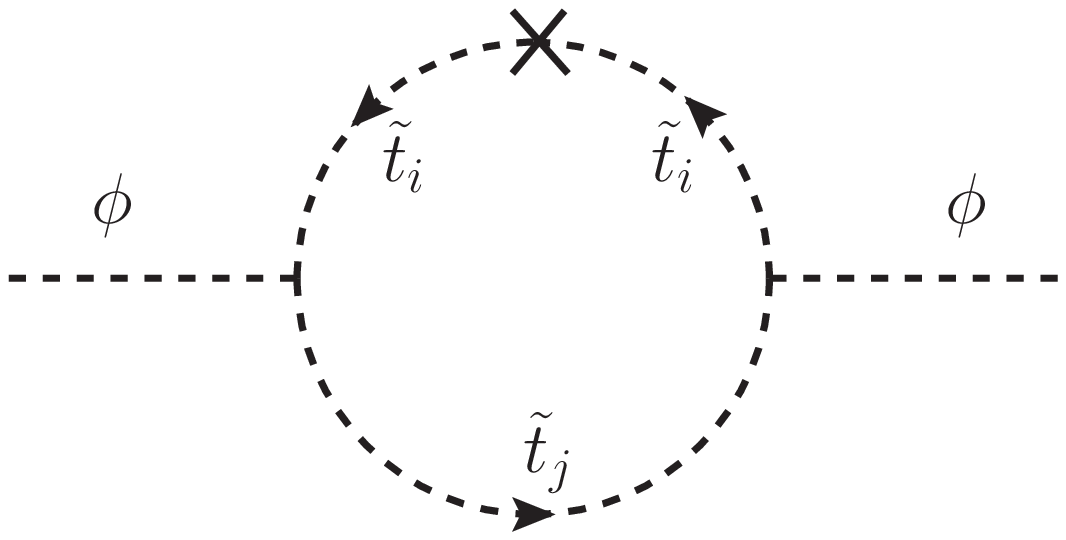}} \hspace{20pt} 
\raisebox{-2pt}{\includegraphics[width=0.2\textwidth]{ct_top6}}  
\caption{One-loop subrenormalization diagrams containing top-squark 
  loops with counterterm insertions.  }
\label{fig:stop1Lct}
\end{center}
\end{figure}

Equivalent momentum-dependent terms of \order{\eps} of the scalar-top 
mass counterterms, evaluated from the diagrams in the lower row of 
\reffi{fig:fd_ctis}, do not contribute.  The diagrams with top-squark 
counterterm insertions are depicted in \reffi{fig:stop1Lct}.  The first 
diagram is momentum independent.  In the second diagram, the 
corresponding loop integral is a massive scalar three-point function 
($C_0$) with only scalar particles running in the loop, and thus is UV 
finite.  Consequently, the top-squark mass counterterm insertions of 
\order{\eps} do not contribute.  In the third diagram the stop mass 
counterterm can enter via the (dependent) counterterm for 
$\At$~\cite{mhiggslong,SbotRen}. This diagram does not possess a 
momentum-dependent divergence, however, and thus the \order{\eps} term 
of the scalar top mass counterterm again does not contribute.

%%%%%%%%%%%%%%%%%%%%%%%%% F I G U R E %%%%%%%%%%%%%%%%%%%%%%%%%%%%%%%%%%%%%%%%%
%\begin{figure}[htb!]
%%%%%%%%%%%%%%%%%%%%%%%%% F I G U R E %%%%%%%%%%%%%%%%%%%%%%%%%%%%%%%%%%%%%%%%%

%%%%%%%%%%%%%%%%%%%%%%%%%%%%%%%%%%%%%%%%%%%%%%%%%%%%%%%%%%%%%%%%%%%%%%%%%%%%%

\subsection{Physics content and interpretation}

In the following we give another view on the finite $\dmteps$ term
from the top mass renormalization and on the interpretation of the
different results for the Higgs-boson masses with and without this term.
 
\bigskip
In the approximation with $p^2=0$ for the two-loop  
self-energies, the results are the same for either dropping or 
including the $\dmteps$ term, provided that this is done
everywhere in the contributions from the top--stop sector  
in the renormalized two-loop self-energies.

\bigskip
As explained above, abandoning the $p^2=0$ approximation
yields an additional $\dmteps$ in the $p^2$-coefficient of the
self-energy $\se{\phi_2}^{(2)} (p^2)$ when the on-shell top-quark
mass counterterm, see Eq.~(\ref{dmt}), is used,
as well as in the $A$-boson self-energy $\se{AA}(p^2)$ from which it 
induces an additive term $\sim M_A^2\, \dmteps/m_t$
to the mass counterterm $\delta M_A^2$. 

In the renormalized  self-energy $\ser{\phi_2}^{(2)} (p^2)$,
Eq.~({\ref{rse22}), 
this extra $p^2$-dependent term survives 
when $\delta Z_{\cHz}^{(2)}$ is defined in the minimal way containing only
the $1/\eps$ and $1/\eps^2$ singular parts; 
however, it disappears in $\ser{\phi_2}^{(2)} (p^2)$ when the minimal
$\delta Z_{\cHz}^{(2)}=\dZ{\cHz}^{\dmtOS\,(2)}\Big|_{\rm div}$
is replaced by
\begin{align}
\dZ{\cHz}^{(2)} \, \to \; 
\dZ{\cHz}^{(2)} - \frac{3\alt}{2\pi} \frac{\dmteps}{\mt} \, ,
\end{align}
which now accommodates also a finite part of two-loop order.

This shift in $\delta Z_{\cHz}^{(2)}$ by a finite term has also an impact on the
counterterm for $\tb$ via $\delta\! \tb = \frac{1}{2} \delta Z_{\cHz}^{(2)} $.
This has the consequence that
  the extra $\dmteps$ term in $\delta M_A^2$ drops out in 
  the constant counterterms 
%$\delta m^{2\, (2)}_{\phi_1}$,
%$\delta m^{2\, (2)}_{\phi_1\phi_2}$,
%$\delta m^{2\, (2)}_{\phi_2} + m^{2\, (2)}_{\phi_2}\, \delta Z_{\cHz}^{(2)} $
for the renormalized self-energies
 $\ser{\phi_{ij}}^{(2)} (p^2)$ 
 in Eq.~(\ref{rMSSM:renses_higgssector})
 because of cancellations with the $\dmteps$ term in
 $\delta \!\tb$ and $\delta Z_{\cHz}^{(2)} $
(this can be seen from the explicit expressions
% given in Ref.~\cite{Mh-p2-BH4}).
given in Eqs.~(\ref{fieldrenconstphi}) and~(\ref{masscounterterms}) ).

Accordingly, keeping or dropping the finite $\dmteps$ part is thus
equivalent to a finite shift in the field-renormalization constant
$\delta Z_{\cHz}$ at the two-loop level, which corresponds to a 
finite shift in $\tb$ as input quantity. 
Numerically, the shift in $\tb$ is small, and cannot explain the differences
in the $M_h$ predictions from the two schemes.
Hence, these differences originate from 
the different $p^2$ coefficients in  $\ser{\phi_2}^{(2)} (p^2)$.

\bigskip
The impact of a modification of the 
two-loop field-renormalization constant 
on the mass $M_h$ can best be studied in terms of the 
self-energy $\se{hh}$ in the $h,H$ basis, which is composed of the
$\se{\phi_{ij}}$ in the following way,
\begin{align}
\se{hh} & = \, \cos^2\!\alpha\, \se{\phi_2}  
                     + \sin^2\!\alpha\, \se{\phi_1}
                      -2\sin\!\alpha \cos\!\alpha \, \se{\phi_1\phi_2} \, ,
\end{align}                     
where only $\se{\phi_2}$ contains the $p^2$-dependent $\dmteps$ 
contribution. In order to simplify the discussion and to point to the main features,
we assume sufficiently large values of $\tb$ that we can write 
%\begin{align}
$\ser{hh}  \simeq \,\ser{\phi_2} $,
%\end{align}
and $h,H$ mixing effects play only a marginal role
(both simplifications apply to the numerical discussions 
in the subsequent section).
Moreover, to simplify the notation, we drop the indices and define
\begin{align}
\se{hh}  \equiv  \se{} , \quad \ser{hh} \equiv \ser{} , \quad 
%\delta Z_{\cHz} \equiv \delta Z \, .
\delta Z_{hh} \equiv \delta Z \, ,
\end{align}
where $\delta Z_{hh} = \cos^2\!\alpha \, \delta Z_{\cHz}
                                  +  \sin^2\!\alpha \, \delta Z_{\cHe}
                                \simeq \delta Z_{\cHz}$.
Starting from the tree-level mass $m_h$ and the renormalized $h$ self-energy
up to the two-loop level,
\begin{align}
\ser{} (p^2)  & = \, \se{} (p^2) -\delta m_h^2 +\delta Z (p^2 - m_h^2) \, ,
\label{renhh}
\end{align}
we obtain the higher-order corrected mass $M_h$ from the pole of the propagator,
{\it i.e.}
\begin{align}
M_h^2 - m_h^2 + \ser{} (M_h^2) & = \, 0 \, .
\label{polehh}
\end{align}
The Taylor-expansion of the unrenormalized self-energy around $p^2=0$, 
\begin{align}
\se{} (p^2) & = \, \se{} (0) + p^2 \, \se{}' (0) + \tilde{\se{}} (p^2) \, ,
\label{Taylorhh}
\end{align}
yields the 
first two terms containing the singularities in $1/\eps$ and 
$1/\eps^2$, and  the residual fully finite and scheme-independent
part denoted by $\tilde{\se{}} (p^2)$.
With this expansion inserted into Eq.~(\ref{renhh})
one obtains from the pole condition Eq.~(\ref{polehh}) the relation
\begin{align}
(M_h^2 - m_h^2) \big[ 1 + \delta Z + \se{}'(0) \big] +
\big[ \se{}(0) - \delta m_h^2  + m_h^2 \, \se{}'(0) \big]
+ \, \tilde{\se{}} (M_h^2) & = \, 0 \, ,
\end{align}
where the expressions in the square brackets are each finite,
irrespective of a possible finite term in the definition of $\delta Z$.  

%\newpage \noindent
\bigskip
Taking into account that $M_h^2$ differs from $m_h^2$ by a 
a higher-order shift, we can replace
\begin{align}
\tilde{\se{}} (M_h^2) &= \, \tilde{\se{}} (m_h^2) 
  + (M_h^2 -m_h^2)\, \tilde{\se{}}' (m_h^2) + \cdots
\end{align}
and obtain 
\begin{align}
M_h^2 - m_h^2 & = \, 
- \frac{\se{}(0) -\delta m_h^2 +m_h^2\, \se{}'(0) + \tilde{\se{}}(m_h^2) }
  {1+\delta Z + \se{}'(0) + \tilde{\se{}}'(m_h^2) } \\[0.3cm] \nonumber
 & = \, - \big[ \se{}(0) -\delta m_h^2 +m_h^2\, \se{}'(0)
                      +\tilde{\se{}}(m_h^2) \big]_{\rm 1loop\,+\,2loop} \\[0.2cm] \nonumber
 & \quad + 
  \big[ \se{}(0) -\delta m_h^2 +m_h^2\, \se{}'(0)  + \tilde{\se{}}(m_h^2)\big]_{\rm 1loop}
  \cdot \big[\delta Z + \se{}'(0) + \tilde{\se{}}'(m_h^2)\big]_{\rm 1loop}  + \, \cdots 
   \end{align}
showing explicitly all terms up to two-loop order. 
It does not contain the two-loop part of the field-renormalization constant,
which indeed would show up at the three-loop level. 
Hence, effects resulting from different conventions for 
$\delta Z^{({\rm 2loop})}$ in the finite part have to be considered in
the current situation as part of the theoretical uncertainty.

%\end{document}

\subsection{Numerical comparison}

In this section the renormalized momentum-dependent \order{p^2 \alt\als} 
self-energy contributions $\De\ser{hh}$, $\De\ser{hH}$, $\De\ser{HH}$ of 
\refeq{eq:DeltaSE} and the mass shifts
\begin{align}
\De \Mh = \Mh - M_{h,0}, \quad \De \MH = \MH - M_{H,0}
\label{dMhH}
\end{align} 
are compared using either $\dmtOS$ or $\dmtFIN$, as discussed above.
$M_{h,0}$ and $M_{H,0}$ denote the Higgs-boson mass predictions 
\emph{without} the newly obtained \order{p^2 \alt\als} corrections.

\medskip

The results are obtained for two different scenarios.  Scenario~1 is 
adopted from the \mhmax\ scenario described
in~\citere{Carena:2013qia}.   
We use the following parameters:
\begin{align}
\mt &= 173.2\gev,\; \msusy=1\tev,\; \Xt =2\,\msusy\; , \non \\
\mgl &= 1500\gev,\; \mu = M_2 = 200\gev\,.
\end{align}
Here $M_2$ denotes the $SU(2)$ soft SUSY-breaking parameter, where
  the $U(1)$ parameter is derived via the GUT relation
$M_1 = (5/3)\, (\sw^2/\cw^2)\, M_2$.
Scenario~2 is an updated version of the ``light-stop scenario'' of
\citeres{Carena:2013qia,updated-Scenarios}
\begin{align}
\mt &= 173.2\gev,\; \msusy= 0.5 \tev,\; \Xt =2\,\msusy\; , \non \\
\mgl &= 1500\gev,\; \mu = M_2 = 400\gev \; M_1 = 340 \gev\,,
\end{align}
leading to stop mass values of 
\begin{align}
\mste = 326.8 \gev, \; \mstz = 673.2 \gev~.
\end{align}
A renormalization scale of $\mu = \mt$ is set in all numerical 
evaluations.

%%%%%%%%%%%%%%%%%%%%%%%%%%%%%%%%%%%%%%%%%%%%%%%%%%%%%%%%%%%%%%%%%%%%%%%%%%%%%%

\subsection*{Self-energies}

\noindent

In \reffi{fig:deltma} we present the results for the $\delta_A$ (upper 
plot) and $\de_{\Si_{22}}$ (lower plot) contributions for $\tb = 5 (20)$ 
in red (blue) in Scenario~1, where $\delta_A,\de_{\Si_{22}}$ are defined 
in \refeqs{eq:deltadef}.  In the upper plot $\de_A(\MA^2)$ ($\de_A(0)$) 
is shown as solid (dashed) line; correspondingly, in the lower plot 
$\de_{\Si_{22}}(p^2)$ ($\de_{\Si_{22}}(0)$) is depicted as solid 
(dashed) line.  The contribution is seen to decrease quadratically with 
$\MA$ or $p\ (:= \sqrt{p^2})$ when including the momentum-dependent 
terms, see \refeq{eq:nocancellation}.  For $\de_A$ it is 
suppressed with $\TQb$.  For high values of $\MA$ and low $\tb$, the 
$\delta_A$ contribution becomes sizable.  Similarly, for large $p$ the 
$\de_{\Si_{22}}$ term becomes sizable, showing the relevance of the 
$\dmteps$ contribution.

%%%%%%%%%%%%%%%%%%%%%%%%% F I G U R E %%%%%%%%%%%%%%%%%%%%%%%%%%%%%%%%%%%%%%%%
\begin{figure}[ht!]
\centering
\includegraphics[width=0.75\textwidth]{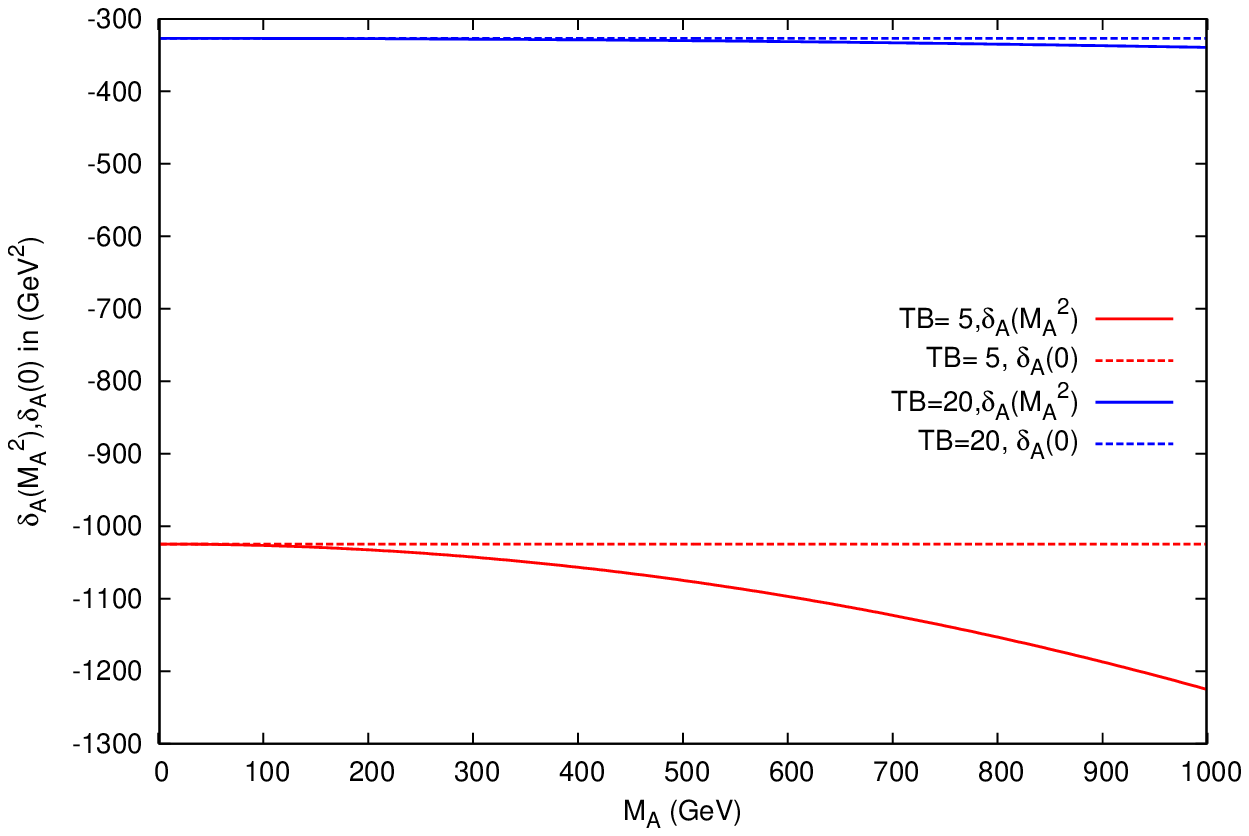}\\
\includegraphics[width=0.75\textwidth]{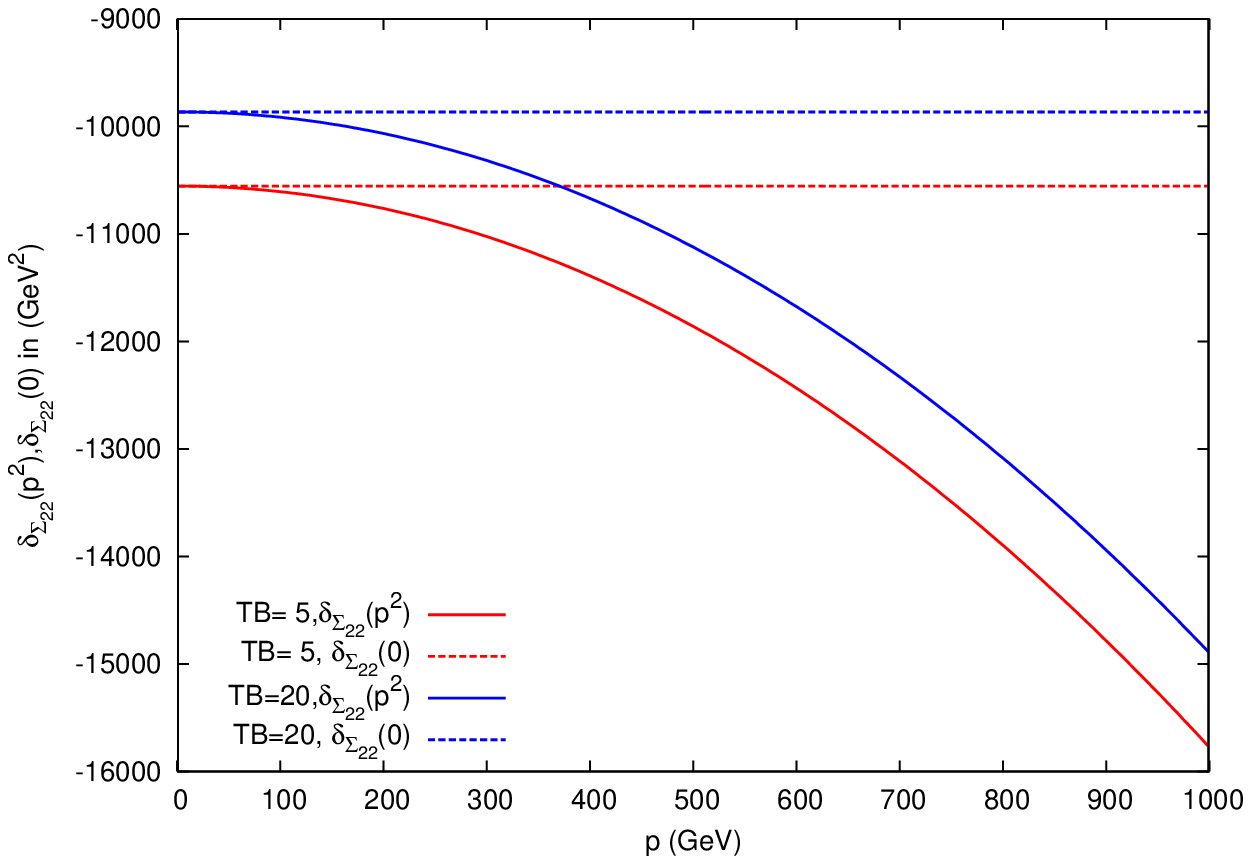}
\caption{
$\delta_{A}(\MA^2)$ and $\delta_{A}(0)$ varying $\MA$ shown in
  the upper plot, $\de_{\Si_{22}}(p^2)$  and $\de_{\Si_{22}}(0)$ in the
  lower plot, both within Scenario 1. 
}
\label{fig:deltma}
\end{figure} 
%%%%%%%%%%%%%%%%%%%%%%%%% F I G U R E %%%%%%%%%%%%%%%%%%%%%%%%%%%%%%%%%%%%%%%%

%%%%%%%%%%%%%%% F I G U R E %%%%%%%%%%%%%%%%%%%%%%%%%%%%%%%%%
\begin{figure}[ht!]
\centering
\includegraphics[width=0.42\textheight]{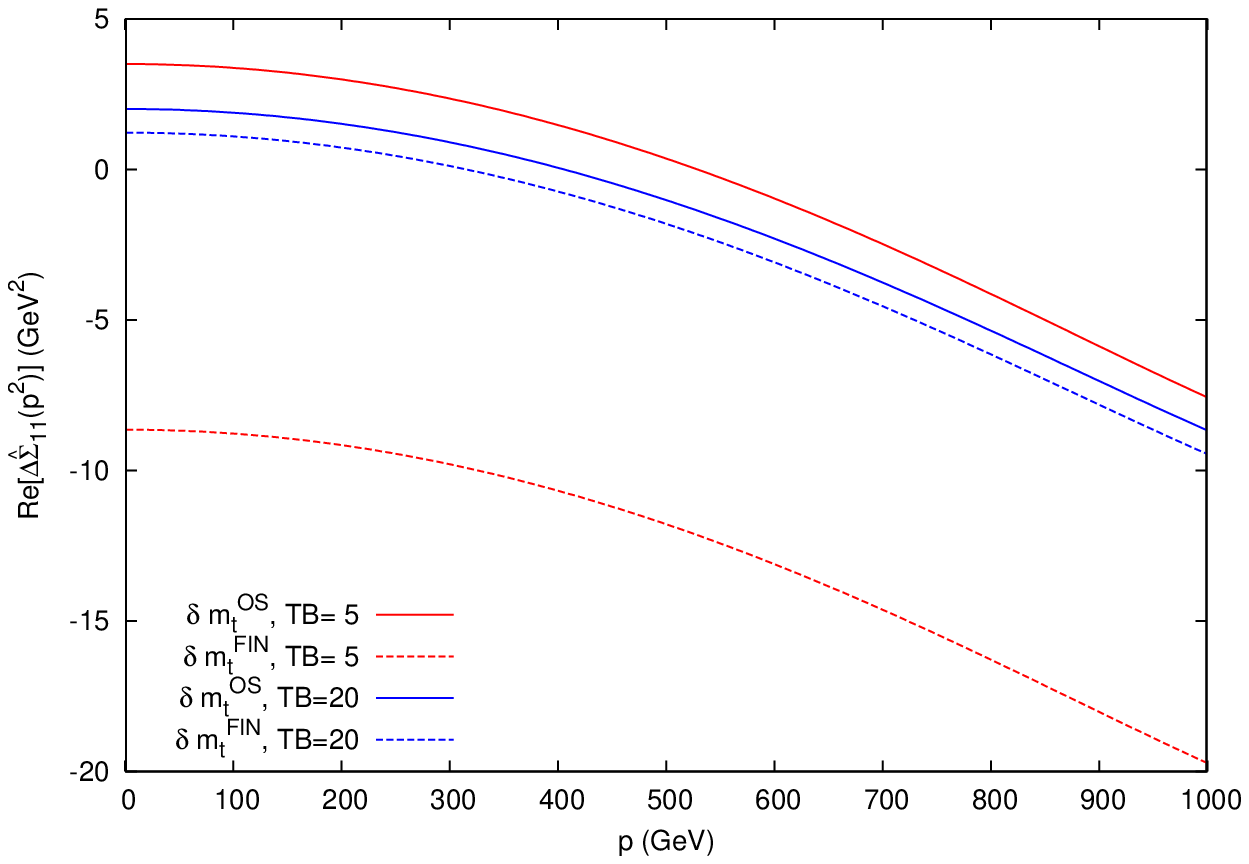}\\
\includegraphics[width=0.42\textheight]{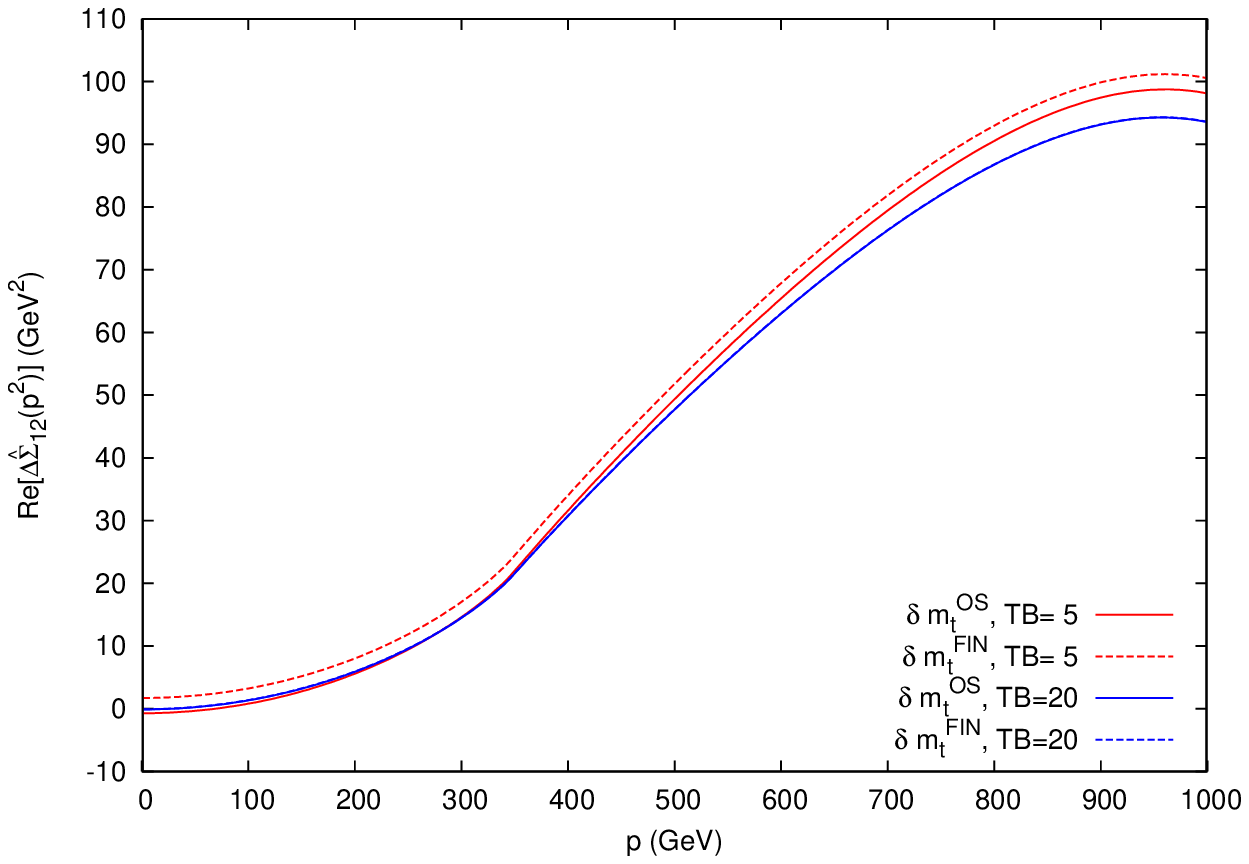}\\
\includegraphics[width=0.42\textheight]{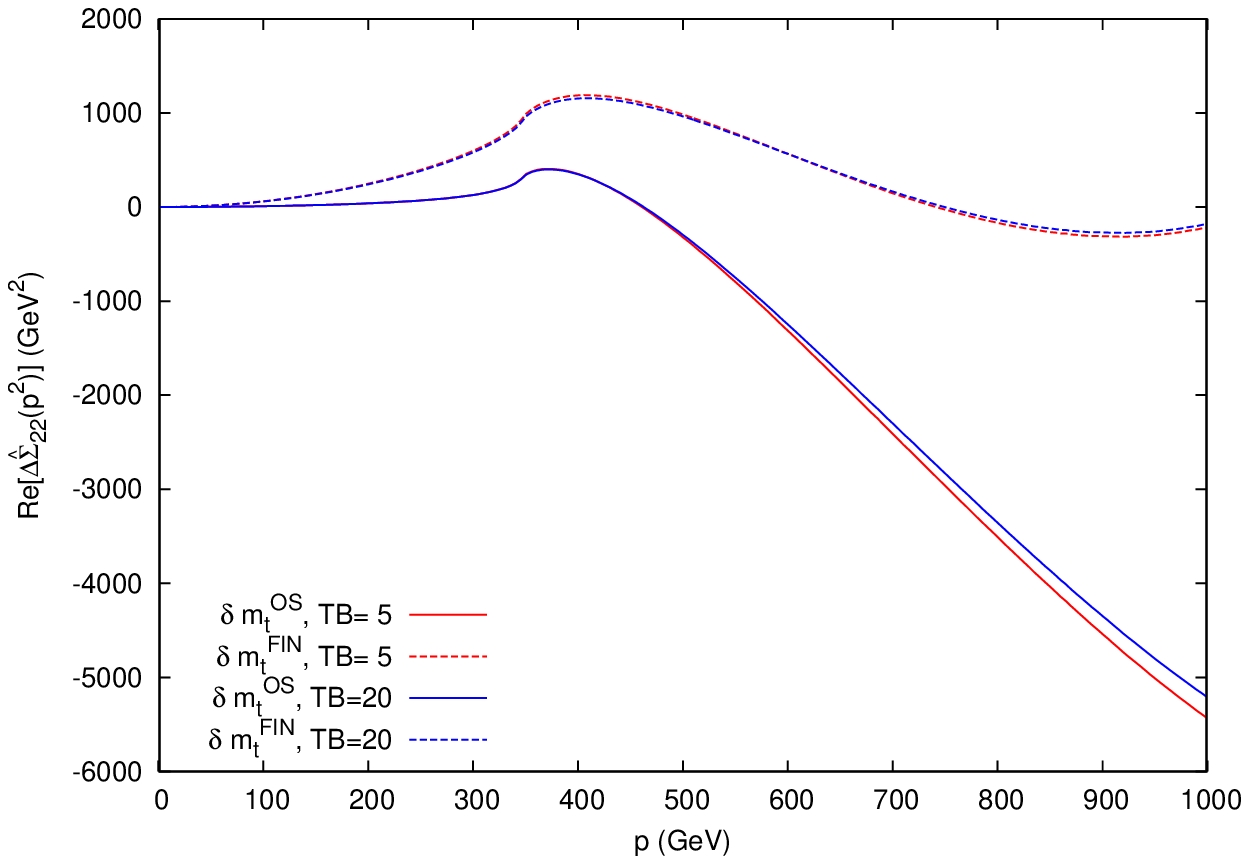}
\caption{$\Delta \hat{\Sigma}_{\phi_{ij}}$ in Scenario~1 (with
$\MA=250 \gev$) for $ij = 11, 12, 22$ in the upper, the
middle and the lower plot, respectively.  The solid (dashed) lines
show the result obtained with $\dmtOS$ ($\dmtFIN$); the red (blue) lines
correspond to $\tb = 5 (20)$.
}
\label{fig:deltasigmahatswithma}
\end{figure} 
%%%%%%%%%%%%%%%%%%%%%%%%% F I G U R E %%%%%%%%%%%%%%%%%%%%%%%%%%%%%%%%%%%%%%%%

\medskip

The behavior of the real parts of the two-loop contributions to the 
self-energies $\De\ser{ab}$ is analyzed in 
\reffi{fig:deltasigmahatswithma}.  Solid lines show the result evaluated 
with $\dmtOS$, as obtained in \citere{Mh-p2-BH4} (\ie the new 
contribution added to the previous \fh\ result in \citere{Mh-p2-BH4}, 
see \refeq{eq:DeltaSE}).  Dashed lines show the result evaluated with 
$\dmtFIN$, as obtained in \citere{Mh-p2-DDVS}.  We show $\MA = 250 \gev$ 
and $\tb = 5 (20)$ as red (blue) lines.  The difference between the 
$\dmtFIN$ and $\dmtOS$ calculations for $\De\ser{\Pe}$ and 
$\De\ser{\Pe\Pz}$ is $p$-independent, as discussed below 
\refeq{eq:nocancellation}, and the difference between the two schemes is 
numerically small.  For $\De\ser{\Pz}$, on the other hand, the 
difference becomes large for large values of $p$.  This self-energy 
contribution is mostly relevant for the light $\cp$-even Higgs-boson, 
however, \ie for $p \sim \Mh$, and thus the \emph{relevant} numerical 
difference remains relatively small (but non-zero) compared to the 
larger differences at large $p$.
For completeness it should be mentioned that the imaginary part is not 
affected by the variation of the top-quark renormalization, as only the 
real parts of the counterterm insertions enter the calculation.

Scenario~2 was omitted as the relevant aspects for the analysis of the 
self-energies using $\dmtOS$ vs.\ $\dmtFIN$ have become sufficiently 
apparent within Scenario~1.

%%%%%%%%%%%%%%%%%%%%%%%%%%%%%%%%%%%%%%%%%%%%%%%%%%%%%%%%%%%%%%%%%%%%%%%%%%%%%%

\subsection*{Mass shifts}

We now turn to the effects on the neutral $\cp$-even Higgs-boson masses 
themselves.  The numerical effects on the two-loop corrections to the 
Higgs-boson masses $M_{h,H}$ are investigated by analyzing the mass 
shifts $\De\Mh$ and $\De\MH$ of \refeq{dMhH}.  The results are shown for 
the two renormalization schemes for the top-quark mass, \ie using 
$\dmtOS$ or $\dmtFIN$.  The color coding is as in 
\reffi{fig:deltasigmahatswithma}.
The results for Scenario~1 are shown in 
\reffi{fig:shiftswithma_compareren_scen1} and are in agreement with 
Figs.~2 and~3 (left) in \citere{Mh-p2-DDVS}, \ie we reproduce the 
results of \citere{Mh-p2-DDVS} using $\dmtFIN$.
The results for Scenario~2 are shown in 
\reffi{fig:shiftswithma_compareren_scen2}.  The results are again in 
agreement with Figs.~2 and~3 (right) in \citere{Mh-p2-DDVS}.  This 
agreement confirms the use of $\dmtFIN$ in \citere{Mh-p2-DDVS}, in 
comparison with $\dmtOS$ used in the evaluation of our results.
For the contribution to $\MH$, peaks can be observed at $\MA = 2 \mste$, 
$\mste + \mstz$, $2 \mstz$, see also \citere{Mh-p2-BH4} and the 
discussion of \reffi{fig:drvsos_alsalt_scen2} below.

Since the the results using $\dmtOS$ and $\dmtFIN$ correspond to 
two different renormalization schemes, their difference should be 
regarded as an indication of missing higher-order momentum-dependent 
corrections.

%%%%%%%%%%%%%%% F I G U R E %%%%%%%%%%%%%%%%%%%%%%%%%%%%%%%%%
\begin{figure}[ht!]
\centering
\includegraphics[width=0.7\textwidth]{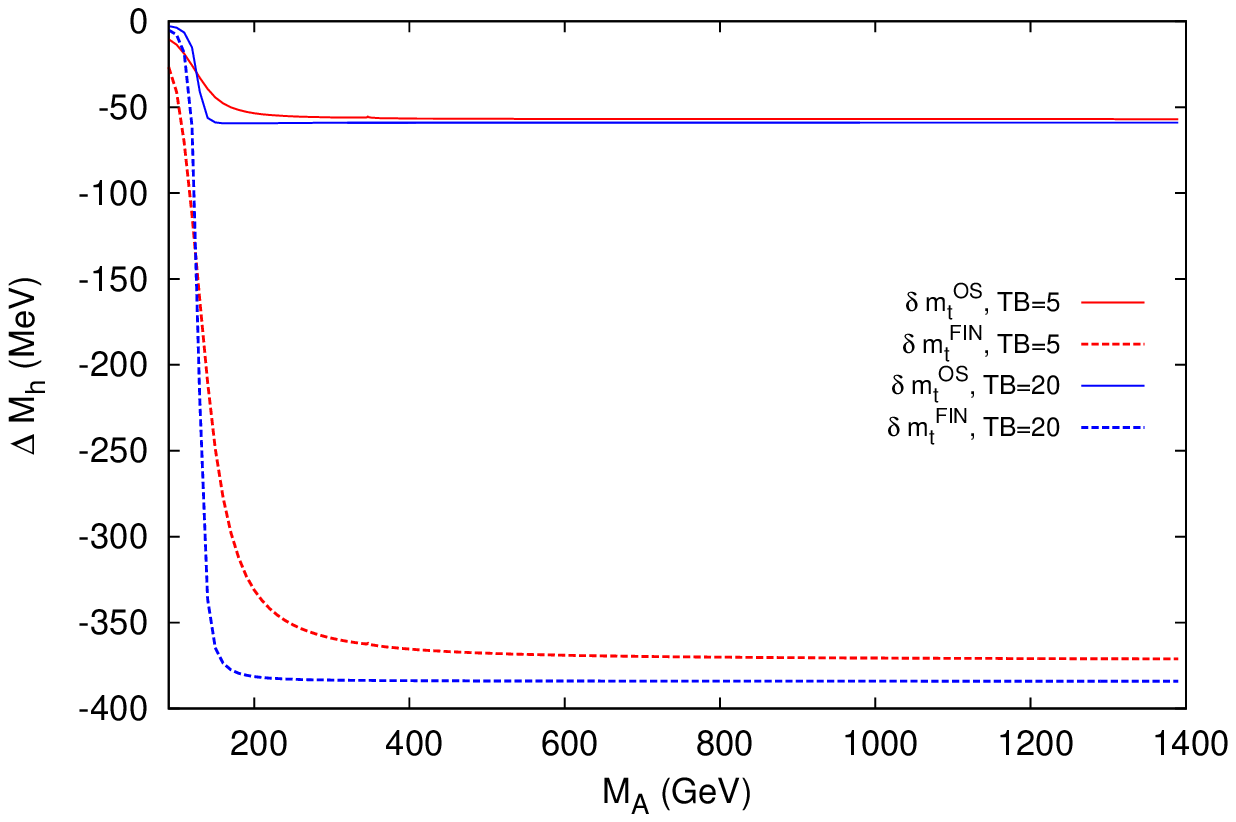}\\
\includegraphics[width=0.7\textwidth]{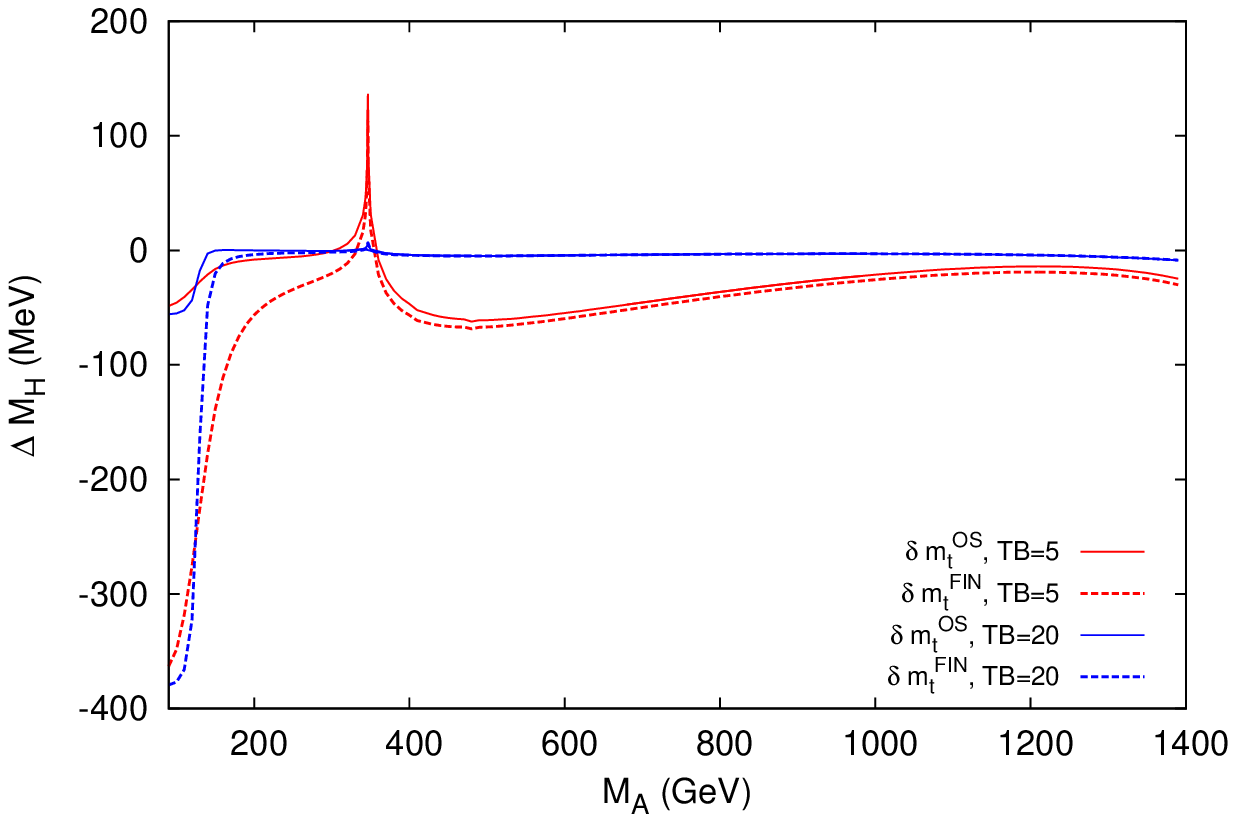}
\caption{Variation of the mass shifts $\Delta\Mh,\Delta\MH$ with the 
$A$-boson mass $\MA$ within Scenario~1, for $\tb=5$ (red) and $\tb = 20$ 
(blue) in- or excluding some $\delta$ terms.  The peak in $\Delta\MH$ 
originates from a threshold at $2\,\mt$.} 
\label{fig:shiftswithma_compareren_scen1}
\end{figure} 
%%%%%%%%%%%%%%% F I G U R E %%%%%%%%%%%%%%%%%%%%%%%%%%%%%%%%%

%%%%%%%%%%%%%%% F I G U R E %%%%%%%%%%%%%%%%%%%%%%%%%%%%%%%%%
\begin{figure}[ht!]
\centering
\includegraphics[width=0.7\textwidth]{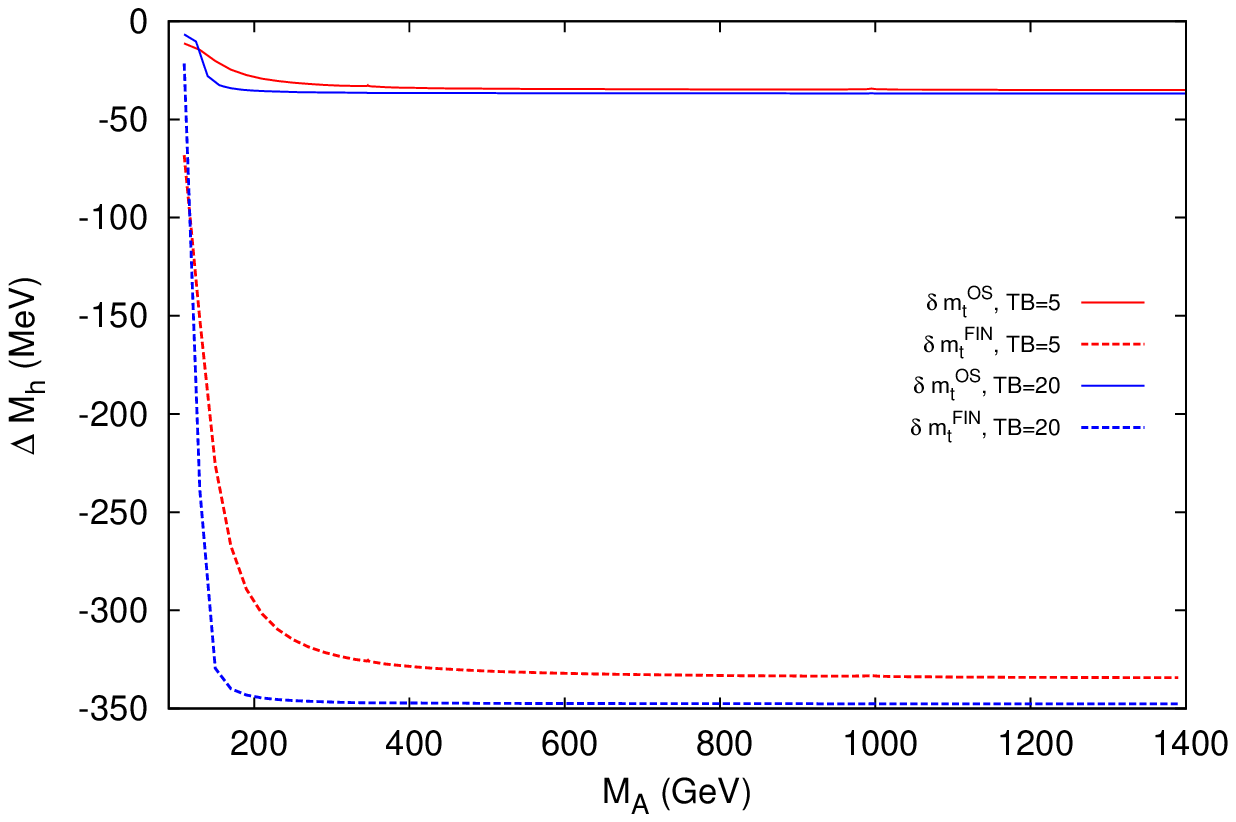}\\
\includegraphics[width=0.7\textwidth]{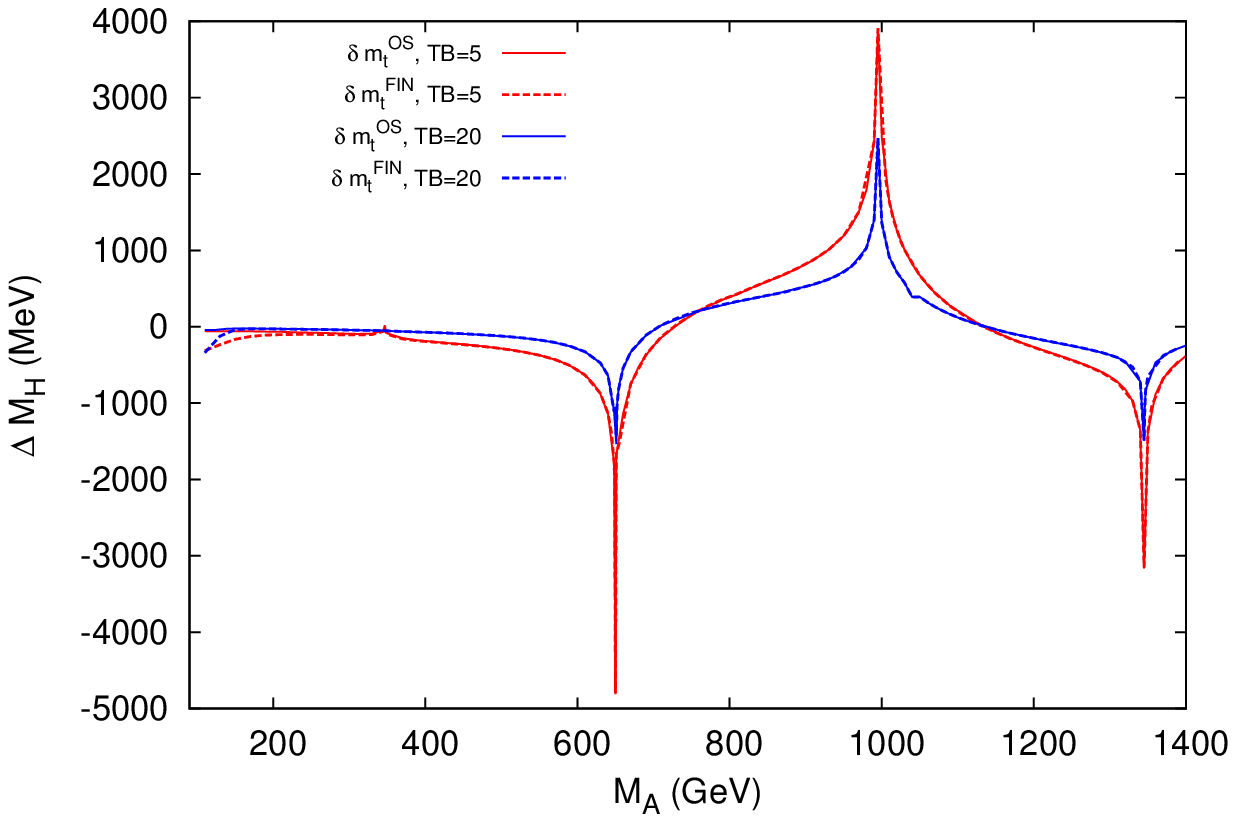}
\caption{Variation of the mass shifts $\Delta\Mh,\Delta\MH$ with the 
$A$-boson mass $\MA$ within Scenario~2, for $\tb=5$ (red) and $\tb = 20$ 
(blue) in- or excluding some $\delta$ terms. The peaks in $\Delta\MH$ 
originate from thresholds at $2\,\mt$, $2\,\mste$, $\mste + \mstz$ and 
$2\,\mstz$, where the threshold at $2\,\mt$ is suppressed by 
$1/\tan^2\be$.} 
\label{fig:shiftswithma_compareren_scen2}
\end{figure} 
%%%%%%%%%%%%%%% F I G U R E %%%%%%%%%%%%%%%%%%%%%%%%%%%%%%%%%

\clearpage

%%%%%%%%%%%%%%%%%%%%%%%%%%%%%%%%%%%%%%%%%%%%%%%%%%%%%%%%%%%%%%%%%%%%%%%%%%
%%%%%%%%%%%%%%%%%%%%%%%%%%%%%%%%%%%%%%%%%%%%%%%%%%%%%%%%%%%%%%%%%%%%%%%%%%

\section{Comparison with the \boldmath{$\mt$ \DRbar} renormalization}
\label{sec:drbar}

Having examined the renormalization of the top-quark mass, we will now 
analyze the numerical differences between an $\mtDRbar$ and an $\mtOS$ 
calculation.  This has been realized by employing a $\DRbar$ 
renormalization of the top-quark mass in all steps of the calculation.  
The top-squark masses are kept renormalized on-shell.  This can be seen 
as an intermediate step towards a full \DRbar\ analysis.

\subsection{Implementation in the program \fh}
\label{sec:fhimp}

In the $\DRbar$ scheme the top-quark mass parameter entering the 
calculation is the MSSM $\DRbar$ top-quark mass, which at one-loop order 
is related to the pole mass $m_t$ (given in the user input) in the 
following way,
\begin{align}
\mtDRbar(\mu) 
&= \mt \cdot \KKL 1 + \frac{\dmtfin}{\mt}
+ \mathcal{O}\KL\bigl(\als^{\DRbar}\bigr)^2\KR \KKR\,.
\label{trafo}
\end{align}
The term $\dmtfin$ can be obtained from \refeq{dmt}, with the formal
replacement $\alpha_s \rightarrow \als^{\DRbar}(\mu)$, yielding
\begin{align}
\nonumber  \frac{\dmtfin}{\mt} = \als^{\DRbar}(\mu) 
& \left(   -\frac{5}{3\pi} + \frac{1}{\pi}\log(\mt^2/\mu^2) 
   + \frac{\mgl^2}{3 \mt^2 \pi} \left(-1 + \log(\mgl^2/\mu^2)\right) \right. \\
\nonumber &   + \frac{1}{6 \mt^2 \pi} \Bigl(\mste^2 (1 - \log(\mste^2/\mu^2)) 
     +  \mstz^2 (1 - \log(\mstz^2/\mu^2)) \\
\nonumber &  + (\mgl^2 + \mt^2 - \mste^2 - 2 \mgl \mt \sin(2 \theta_t))
 \re[B_0^{\rm fin}(\mt^2, \mgl^2, \mste^2)]\\
&  + (\mgl^2 + \mt^2 - \mstz^2 + 2 \mgl \mt \sin(2 \theta_t))
\re[B_0^\mathrm{fin}(\mt^2, \mgl^2, \mstz^2)] \Bigr) \biggr).
\end{align}
At zeroth order, 
$\als^{\DRbar}(\mu)=\als^{\MSbar}(\mu)$. 
%We choose $\mu=m_t$ as renormalization scale.

As on-shell renormalized quantities the stop masses $\mste$ and $\mstz$ 
should have fixed values, independently of the renormalization chosen 
for the top-quark mass.  We compensate for the changes induced by 
$\dmtfin$ in the stop mass matrix, \refeq{Sfermionmassenmatrix}, by 
shifting the SUSY-breaking parameters as follows,
\begin{subequations}
\begin{align}
\MstL^2 &\to \MstL'^2 = \MstL^2 + (\mtOS)^2 - (\mtDRbar)^2 \,,\\
\MstR^2 &\to \MstR'^2 = \MstR^2 + (\mtOS)^2 - (\mtDRbar)^2 \,,\\
\At &\to \At' = \frac{\mtOS}{\mtDRbar} \KL\At - \frac{\mu}{\tb}\KR +
  \frac{\mu}{\tb}\,.
\end{align}
\end{subequations}
(Except for $\At$, which actually appears in the Feynman rules, \fh\ 
only pretends to perform these shifts but computes the sfermion masses 
using $\mtOS$.)

\medskip

This procedure is available in \fh\ from version 2.11.1 on and is 
activated by setting the new value 2 for the \texttt{runningMT} flag. 
The comparison of the results with \DRbar\ and with OS renormalization 
admits an improved estimate of (some) of the missing three-loop 
corrections in the top/stop sector.

%%%%%%%%%%%%%%%%%%%%%%%%%%%%%%%%%%%%%%%%%%%%%%%%%%%%%%%%%%%%%%%%%%%%%%%%%%%%%%%

\subsection{Numerical analysis}
\label{sec:drbaranalysis}

In the following plots we show the difference
\begin{equation}
\bar\De\Mphi := \Mphi(\mtOS) - \Mphi(\mtDRbar),\quad \phi = h, H,
\end{equation}
between $\Mphi$ evaluated in the OS scheme, \ie using $\mtOS$ 
(\emph{not} $\mtFIN$), and in the \DRbar\ scheme, \ie using $\mtDRbar$.

\subsection*{Dependence on \boldmath{$\MA$}}

In the upper half of \reffi{fig:drvsos_alsalt_scen1}, $\bar\De\Mh$ is 
plotted in Scenario~1 as a function of $\MA$ with $\tb = 5 (20)$ in red 
(blue).  The solid (dashed) lines show the difference evaluated at the 
full one-loop level (including the \order{\alt\als} corrections).  The 
dotted lines include the newly calculated \order{p^2 \alt\als} 
corrections.  For $\MA \gsim 200 \gev$ one observes large differences of 
\order{10 \gev} at the one-loop level, indicating the size of missing 
higher-order corrections from the top/stop sector beyond one-loop.  This 
difference is strongly reduced at the two-loop level, to about $\sim 3 
\gev$, now corresponding to missing higher orders beyond two-loop from 
the top/stop sector.  The dotted lines are barely visible below the 
dashed lines, indicating the relatively small effect of the \order{p^2 
\alt\als} corrections as derived in \citere{Mh-p2-BH4}.

The lower plot of \reffi{fig:drvsos_alsalt_scen1} shows the 
corresponding results for $\bar\De\MH$ with the same color/line coding.  
Here large effects are only visible for low $\MA$, where the 
higher-order corrections to $\MH$ are sizable (and the light Higgs-boson 
receives only very small higher-order corrections).  In this part of the 
parameter space the same reduction of $\bar\De\MH$ going from one-loop 
to two-loop can be observed.

The behavior is similar for Scenario~2, shown in 
\reffi{fig:drvsos_alsalt_scen2} (with the same line/color coding as in 
\reffi{fig:drvsos_alsalt_scen1}), only the size of the difference 
$\bar\De\Mh$ is $\sim 20\%$ smaller at the one-loop, and $\sim 50\%$ 
smaller at the two-loop level compared to Scenario~1.  The same peak 
structure due to thresholds as in 
\reffi{fig:shiftswithma_compareren_scen2} is visible.

%%%%%%%%%%%%%%% F I G U R E %%%%%%%%%%%%%%%%%%%%%%%%%%%%%%%%%
\begin{figure}[ht!]
\centering
\includegraphics[width=0.7\textwidth]{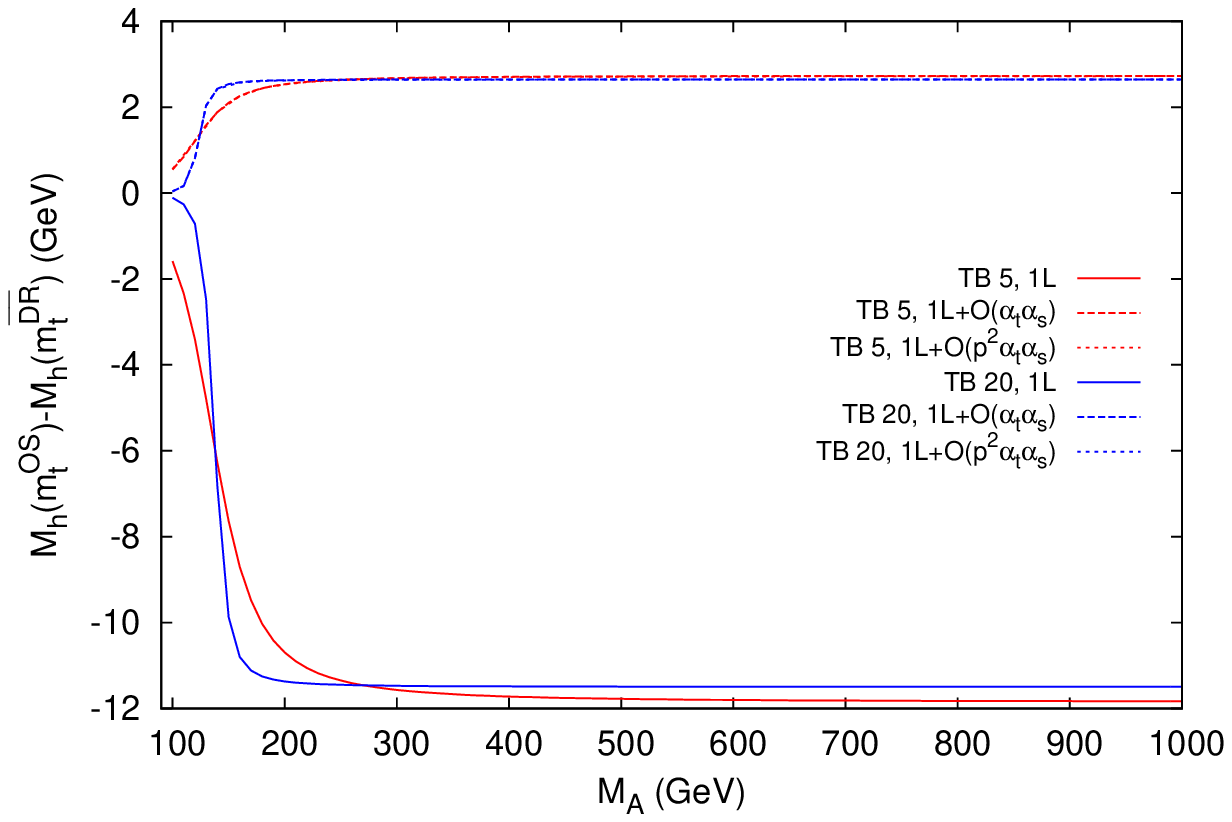}\\
\includegraphics[width=0.7\textwidth]{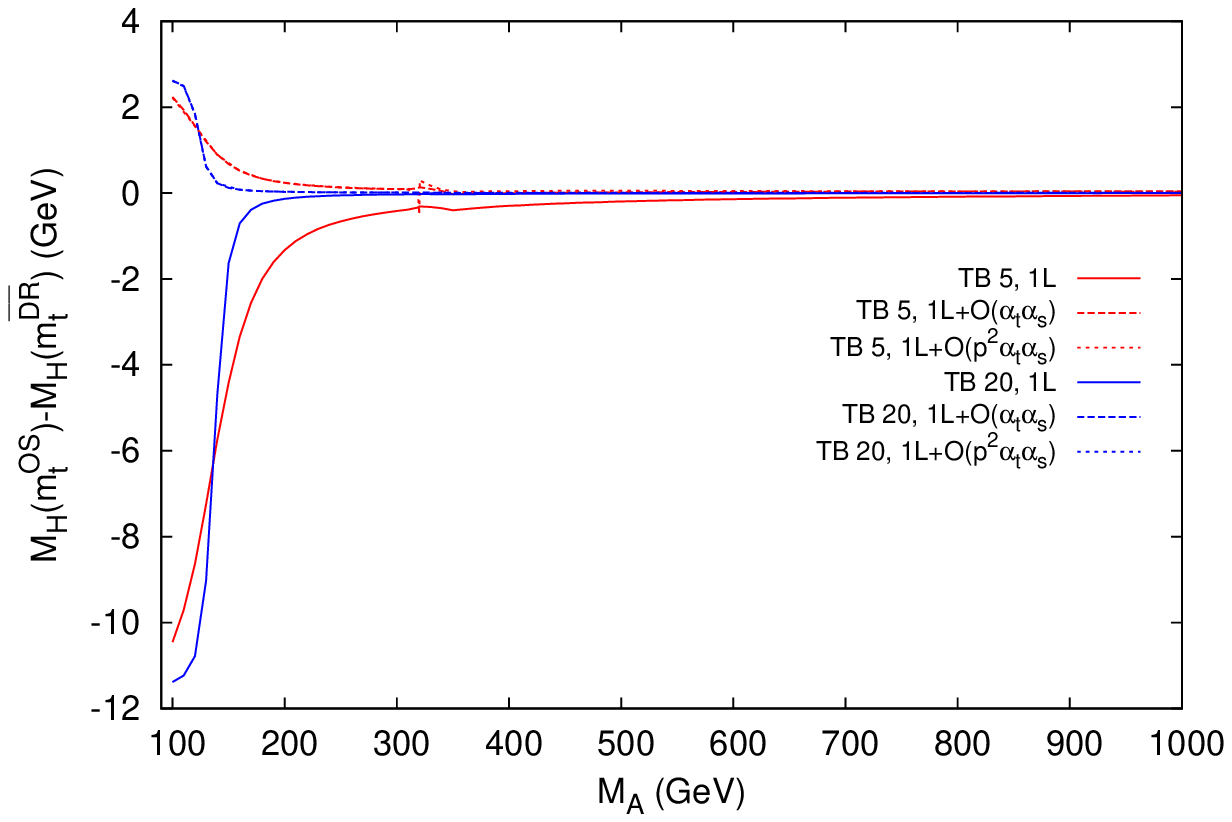}
\caption{$\bar\De\Mphi = \Mphi(\mtOS) - \Mphi(\mtDRbar)$ for $\phi = h$ 
(upper plot) and $\phi = H$ (lower plot).  The difference is shown as 
solid (dashed/dotted) line at the one-loop 
(\order{\alt\als}/\order{p^2\alt\als}) level as a function of $\MA$ for 
$\tb = 5 (20)$ in red (blue) within Scenario~1.}
\label{fig:drvsos_alsalt_scen1}
\end{figure} 
%%%%%%%%%%%%%%% F I G U R E %%%%%%%%%%%%%%%%%%%%%%%%%%%%%%%%%

%%%%%%%%%%%%%%% F I G U R E %%%%%%%%%%%%%%%%%%%%%%%%%%%%%%%%%
\begin{figure}[ht!]
\centering
\includegraphics[width=0.7\textwidth]{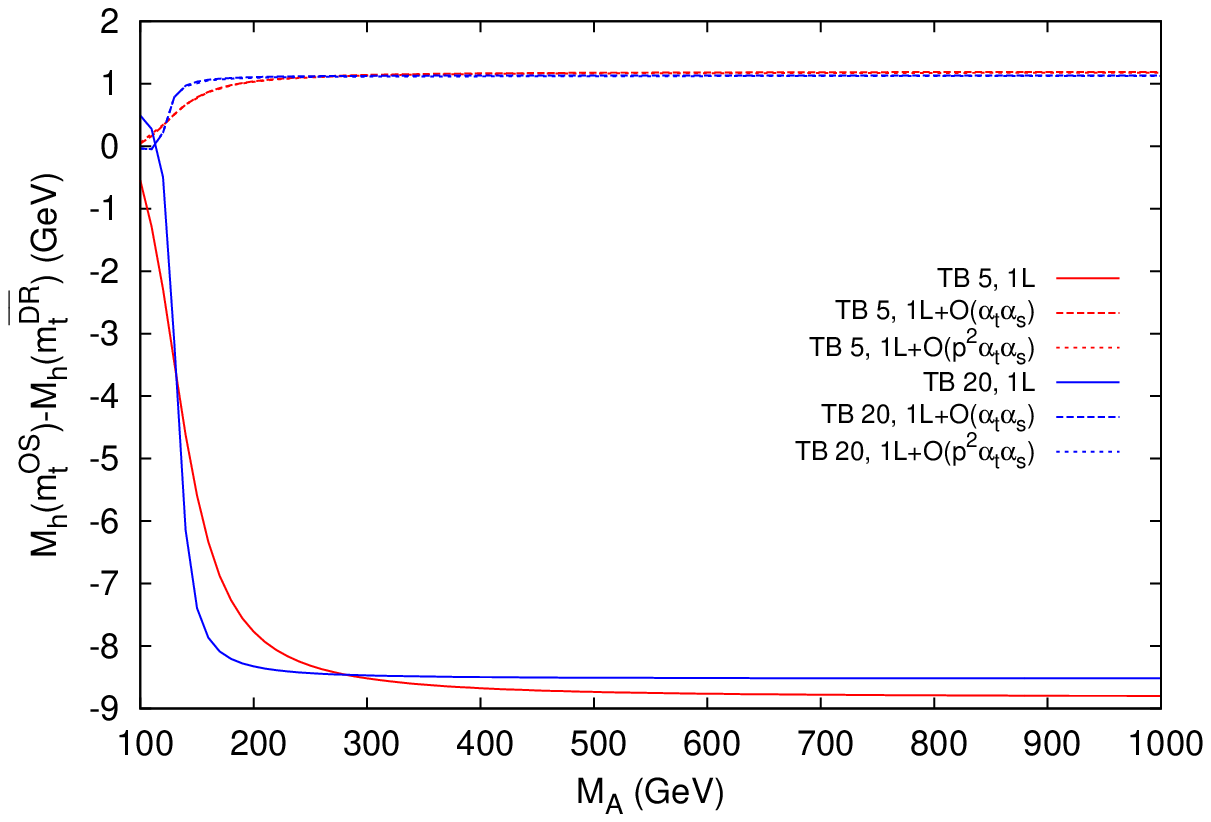}\\
\includegraphics[width=0.7\textwidth]{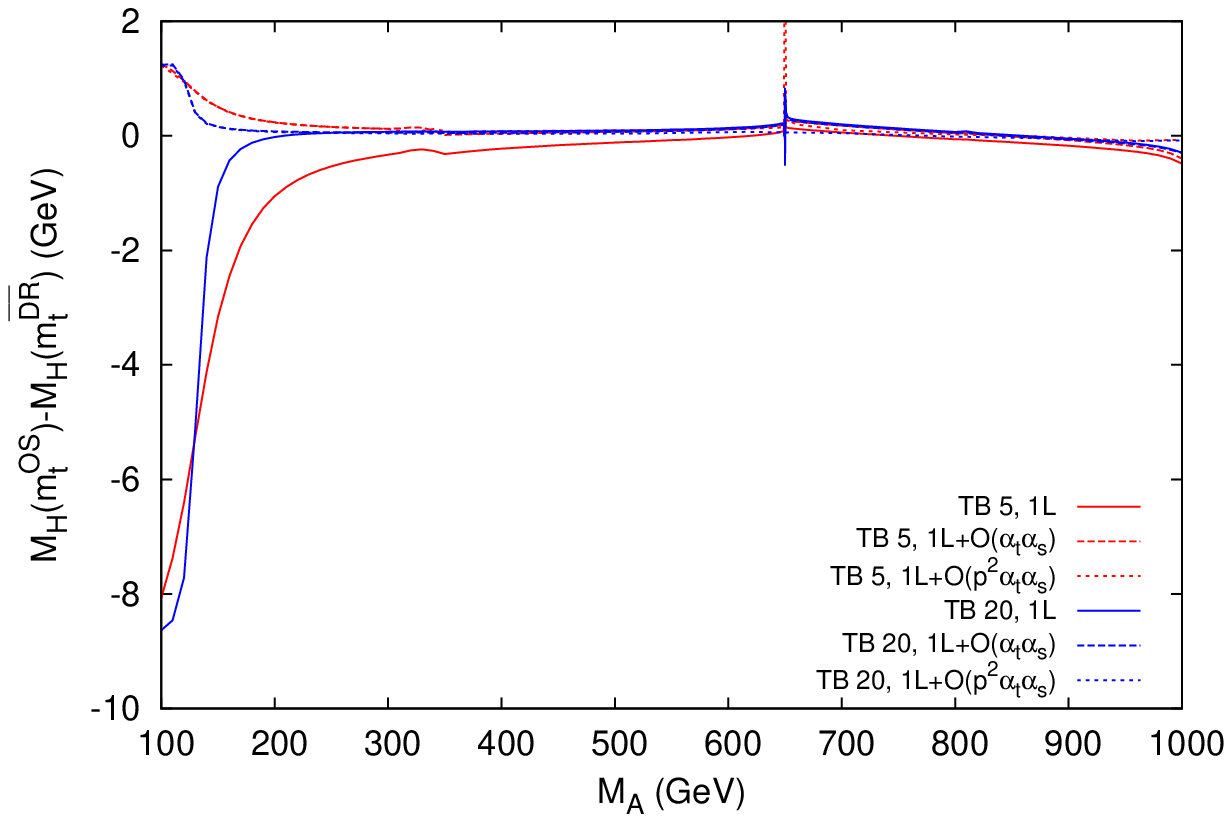}
\caption{$\bar\De\Mphi = \Mphi(\mtOS) - \Mphi(\mtDRbar)$ for $\phi = h$ 
(upper plot) and $\phi = H$ (lower plot) as a function of $\MA$ within 
Scenario~2, with the same line/color coding as in 
\reffi{fig:drvsos_alsalt_scen1}.  The peak in the lower plot originates 
from a threshold at $2\,\mste$.  The threshold at $2\,\mt$ is suppressed 
by $1/\tan^2\beta$.}
\label{fig:drvsos_alsalt_scen2}
\end{figure} 
%%%%%%%%%%%%%%% F I G U R E %%%%%%%%%%%%%%%%%%%%%%%%%%%%%%%%%

\clearpage

\subsection*{Dependence on \boldmath{$\mgl$}}

In \reffis{fig:drvsos_alsalt_mgl_scen1}, 
\ref{fig:drvsos_alsalt_mgl_scen2} we analyze $\bar\De\Mphi$ as a 
function of $\mgl$ in Scenario~1 and~2, respectively.  We fix $\MA = 250 
\gev$ and use the same line/color coding as in 
\reffi{fig:drvsos_alsalt_scen1}. Due to the choice of an MSSM $\DRbar$ 
top-quark mass definition, $\mtDRbar$ varies with $\mgl$ already at the 
one-loop level.

In the upper plots we show the light $\cp$-even Higgs-boson case, where 
it can be observed that the scheme dependence is strongly reduced at the 
two-loop level.  It reaches $2-3 \gev$ in Scenario~1 and $\sim 1 \gev$ 
in Scenario~2, largely independently of $\tb$.  At the one-loop level 
the scheme dependence grows with $\mgl$, whereas the dependence is much 
milder at the two-loop level.  The effects of the \order{p^2\alt\als} 
corrections become visible at larger $\mgl$, in agreement with 
\citere{Mh-p2-BH4}.

The heavy $\cp$-even Higgs-boson case is shown in the lower plots.  At 
small $\tb$ scheme differences of \order{600 \mev (150 \mev)} can be 
observed at the one-(two-)loop level.  For large $\tb$ the differences 
always stay below \order{50 \mev}, in agreement with 
\reffi{fig:drvsos_alsalt_scen1}.  The dependence on $\mgl$ is similar as 
for the light Higgs-boson, but again somewhat weaker.

%%%%%%%%%%%%%%% F I G U R E %%%%%%%%%%%%%%%%%%%%%%%%%%%%%%%%%
\begin{figure}[ht!]
\centering
\includegraphics[width=0.7\textwidth]{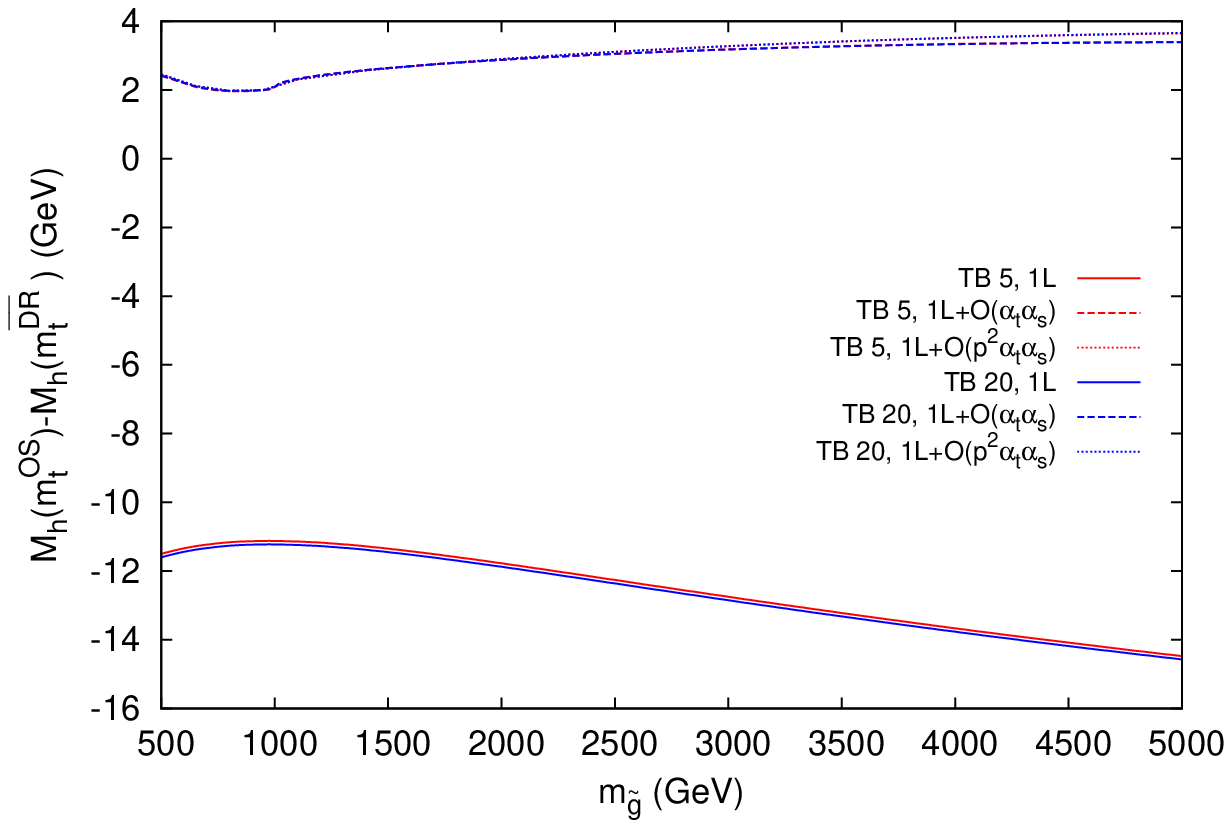}\\
\includegraphics[width=0.7\textwidth]{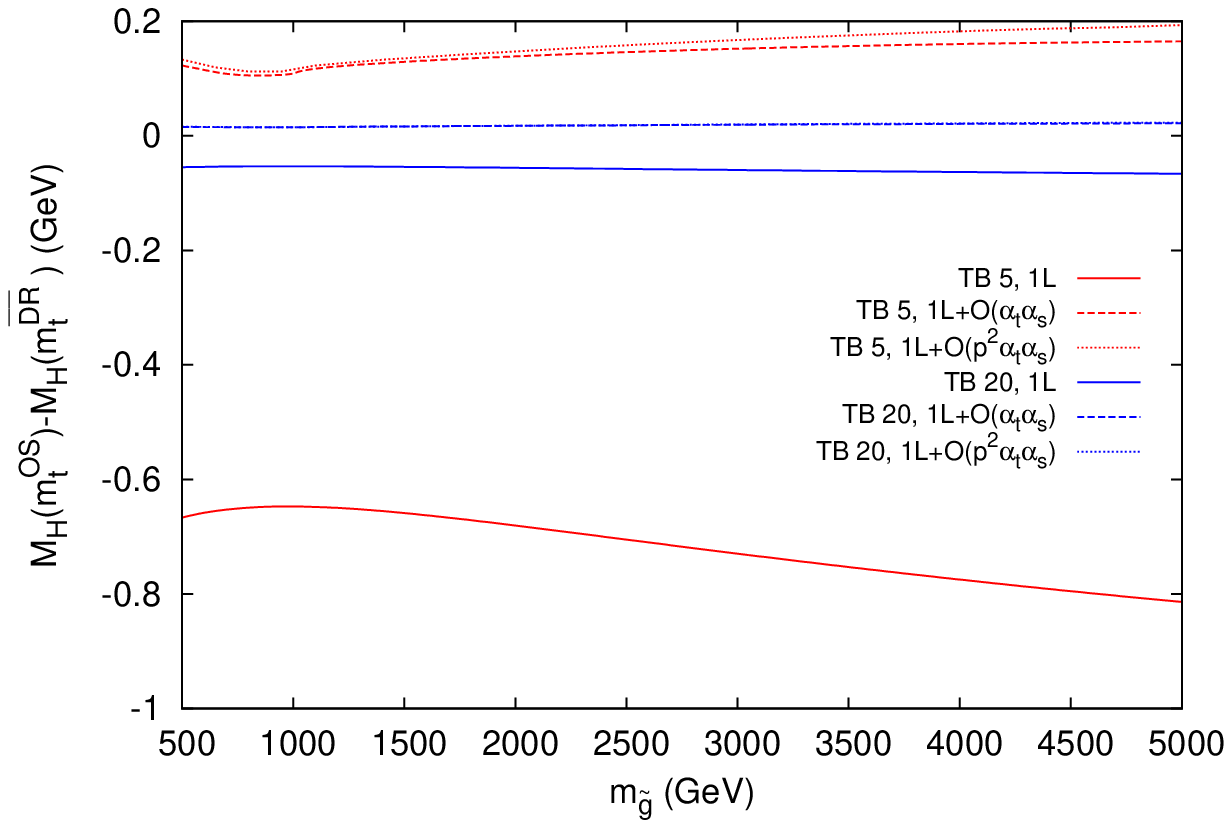}
\caption{$\bar\De\Mphi = \Mphi(\mtOS) - \Mphi(\mtDRbar)$ for $\phi = h$ 
(upper plot) and $\phi = H$ (lower plot) as a function of $\mgl$ within 
Scenario~1, for $\MA = 250 \gev$ and with the same line/color coding as 
in \reffi{fig:drvsos_alsalt_scen1}.}
\label{fig:drvsos_alsalt_mgl_scen1}
\end{figure} 
%%%%%%%%%%%%%%% F I G U R E %%%%%%%%%%%%%%%%%%%%%%%%%%%%%%%%%

%%%%%%%%%%%%%%% F I G U R E %%%%%%%%%%%%%%%%%%%%%%%%%%%%%%%%%
\begin{figure}[ht!]
\centering
\includegraphics[width=0.7\textwidth]{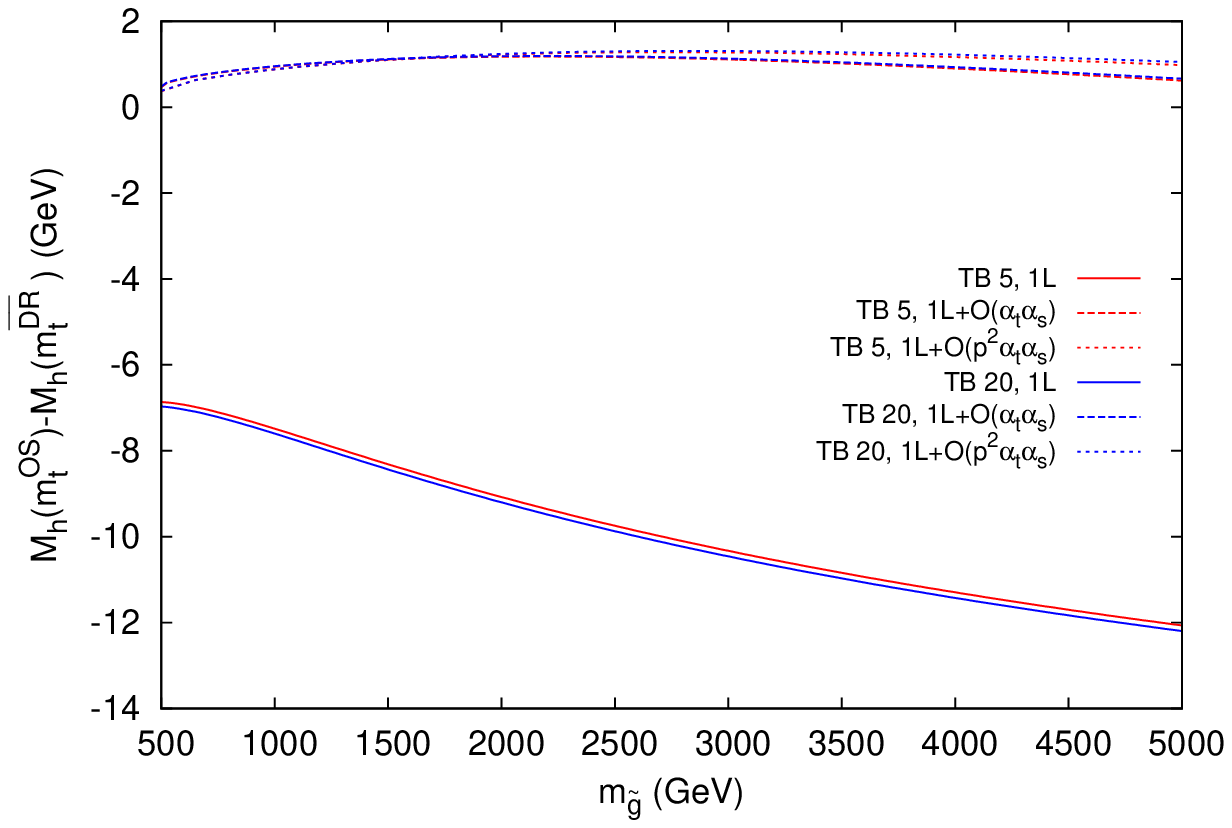}\\
\includegraphics[width=0.7\textwidth]{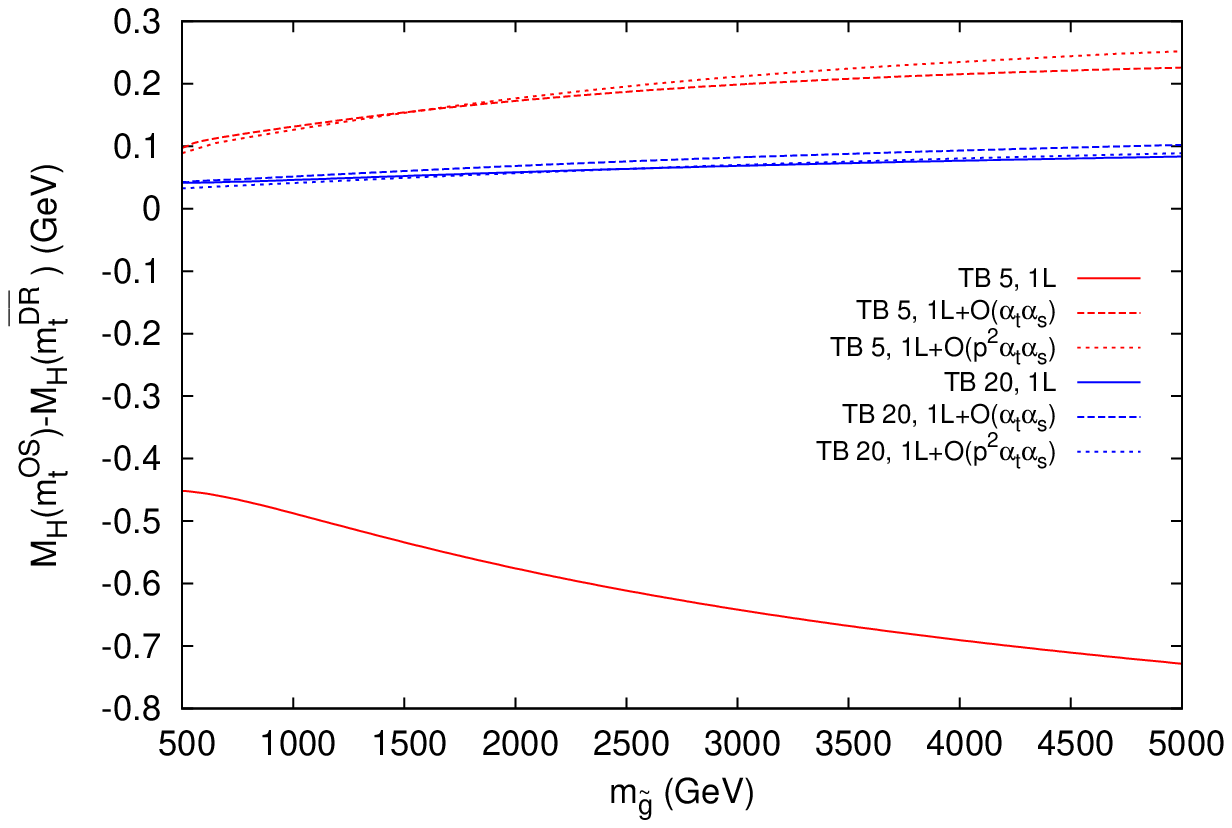}
\caption{$\bar\De\Mphi = \Mphi(\mtOS) - \Mphi(\mtDRbar)$ for $\phi = h$ 
(upper plot) and $\phi = H$ (lower plot) as a function of $\mgl$ within 
Scenario~2, for $\MA = 250 \gev$ and with the same line/color coding as 
in \reffi{fig:drvsos_alsalt_scen1}.}
\label{fig:drvsos_alsalt_mgl_scen2}
\end{figure} 
%%%%%%%%%%%%%%% F I G U R E %%%%%%%%%%%%%%%%%%%%%%%%%%%%%%%%%

\subsection*{Dependence on \boldmath{$\Xt$}}

Finally, in \reffis{fig:drvsos_alsalt_xt_scen1}, 
\ref{fig:drvsos_alsalt_xt_scen2} we analyze $\bar\De\Mphi$ as a function 
of $\Xt = \Xt^{\OS}$ in Scenario~1 and~2, respectively.  We again fix 
$\MA = 250 \gev$ and use the same line/color coding as in 
\reffi{fig:drvsos_alsalt_scen1}.

In the upper plots we show the light $\cp$-even Higgs-boson case.  As 
before the scheme dependence is strongly reduced when going from the 
one-loop to the two-loop case.  In general a smaller scheme dependence 
is found from small $\Xt$, while it increases for larger $|\Xt|$ values, 
in agreement with \citere{mhiggsWN}.  For most parts of the parameter 
space, when the two-loop corrections are included, 
it is found to be below $\sim 3 \gev$. 
The contribution of 
\order{p^2\alt\als} remains small for all $\Xt$ values.

In the heavy $\cp$-even Higgs-boson case, shown in the lower plots, the 
dependence of the size of the effects is slightly more involved, though 
the general picture of a strongly reduced scheme dependence can be 
observed here, too.  
In both scenarios, for large negative $\Xt$ and $\tb = 5$ the 
\order{p^2 \alt \als} contributions can become sizable with respect to
the \order{\alt\als} corrections.

\medskip

In conclusion, the scheme dependence is found to be reduced 
substantially when going from the pure one-loop calculation to the 
two-loop \order{\alt\als} corrections. This indicates that corrections 
at the three-loop level and beyond, stemming from the top/stop sector 
are expected at the order of the observed scheme dependence, \ie at the 
level of $\sim 3 \gev$. This is in agreement with existing calculations 
beyond two-loop~\cite{Mh-logresum,mhiggsFD3l}.
A further reduction of the scheme dependence might be expected by adding 
the \order{\alt^2} contributions.  The $\mtDRbar$ value calculated at 
\order{\als + \alt} is substantially closer to $\mtOS$, reducing already 
strongly the scheme dependence at the one-loop level.  This extended 
analysis is beyond the scope of our paper, however.

%%%%%%%%%%%%%%% F I G U R E %%%%%%%%%%%%%%%%%%%%%%%%%%%%%%%%%
\begin{figure}[ht!]
\centering
\includegraphics[width=0.7\textwidth]{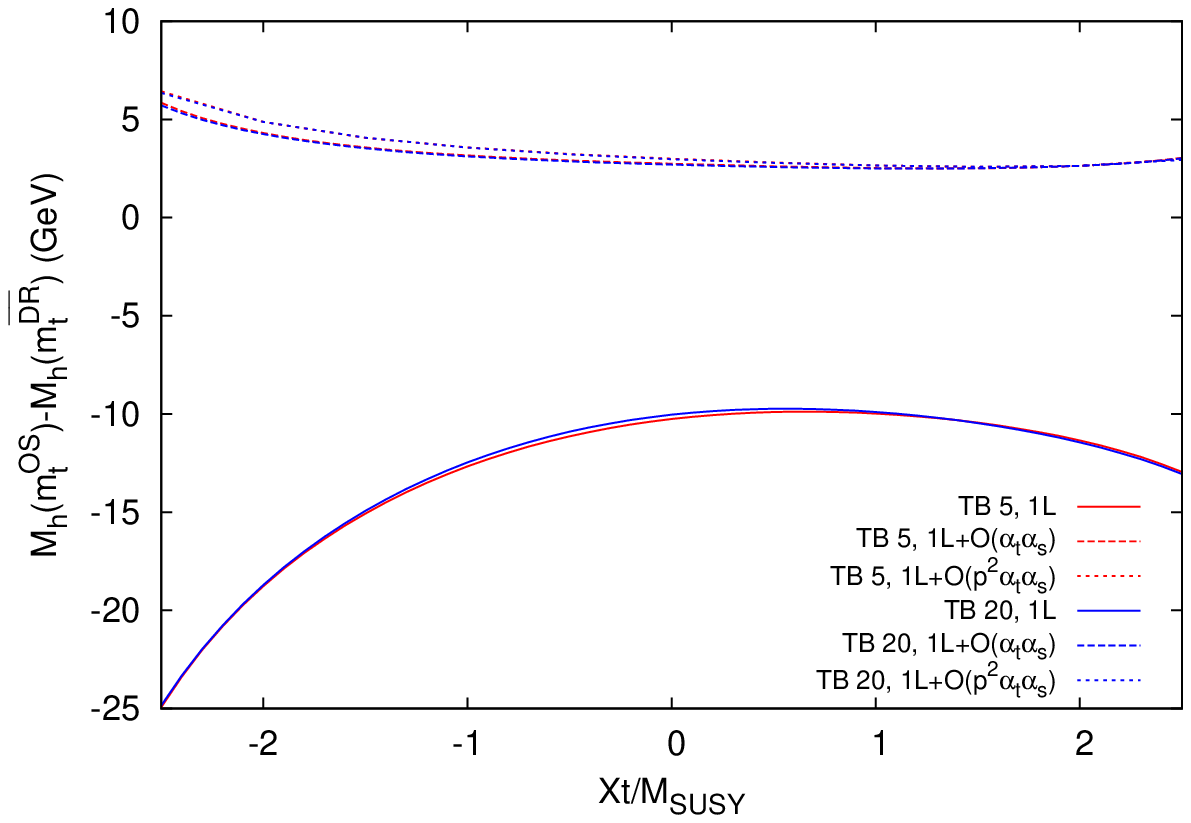}\\
\includegraphics[width=0.7\textwidth]{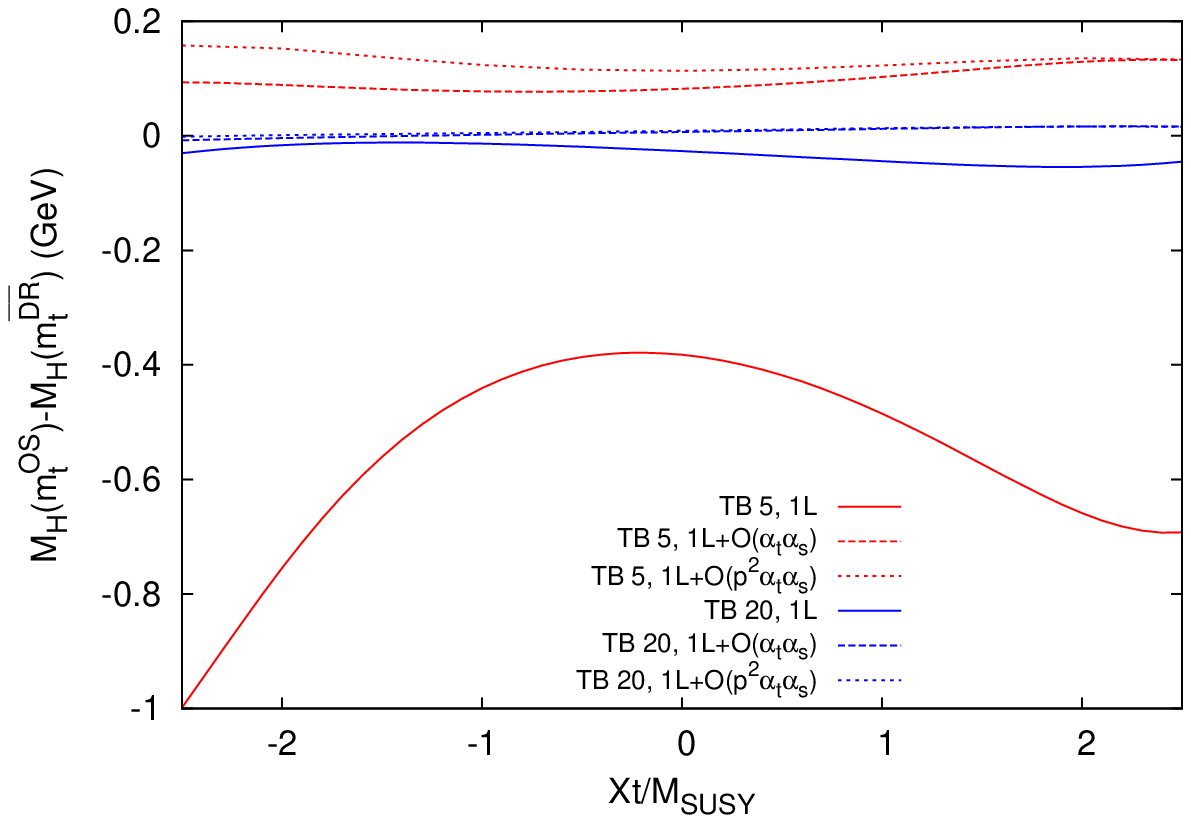}
\caption{$\bar\De\Mphi = \Mphi(\mtOS) - \Mphi(\mtDRbar)$ for $\phi = h$ 
(upper plot) and $\phi = H$ (lower plot) as a function of $\Xt = 
\Xt^{\OS}$ within Scenario~1, with the same line/color coding as in 
\reffi{fig:drvsos_alsalt_scen1}.} 
\label{fig:drvsos_alsalt_xt_scen1}
\end{figure} 
%%%%%%%%%%%%%%% F I G U R E %%%%%%%%%%%%%%%%%%%%%%%%%%%%%%%%%

%%%%%%%%%%%%%%% F I G U R E %%%%%%%%%%%%%%%%%%%%%%%%%%%%%%%%%
\begin{figure}[ht!]
\centering
\includegraphics[width=0.7\textwidth]{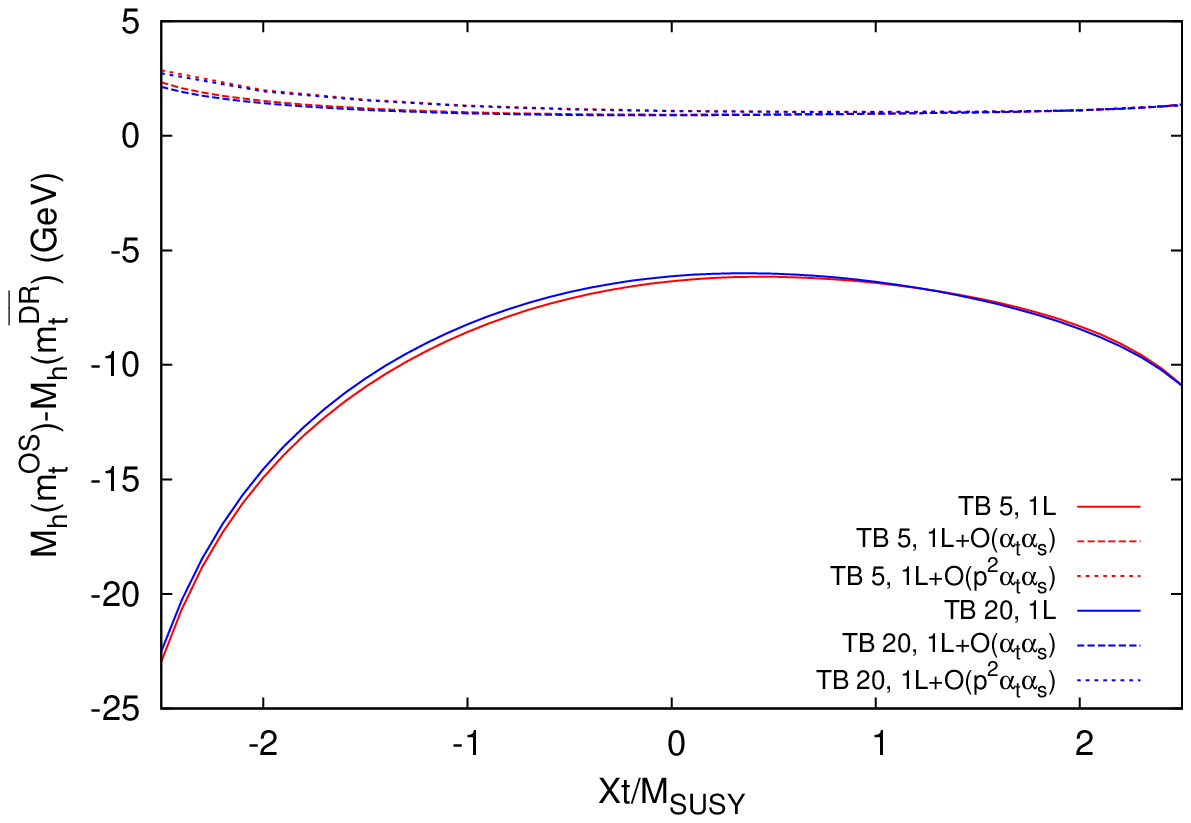}\\
\includegraphics[width=0.7\textwidth]{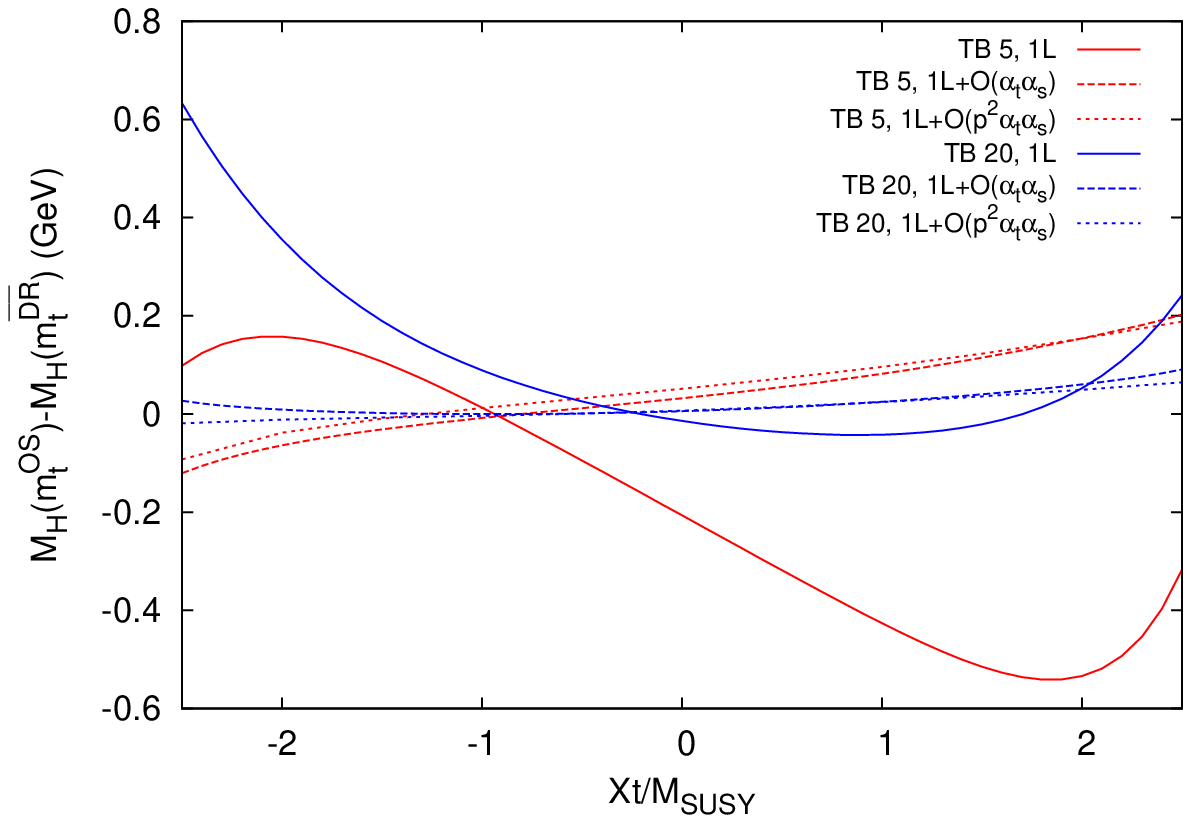}
\caption{$\bar\De\Mphi = \Mphi(\mtOS) - \Mphi(\mtDRbar)$ for $\phi = h$ 
(upper plot) and $\phi = H$ (lower plot) as a function of $\Xt = 
\Xt^{\OS}$ within Scenario~2, with the same line/color coding as in 
\reffi{fig:drvsos_alsalt_scen1}.}
\label{fig:drvsos_alsalt_xt_scen2}
\end{figure} 
%%%%%%%%%%%%%%% F I G U R E %%%%%%%%%%%%%%%%%%%%%%%%%%%%%%%%%

\clearpage 

\section{Conclusions}
\label{sec:conclusions}

In this paper we analyzed the scheme dependence of the \order{\alt\als} 
corrections to the neutral $\cp$-even Higgs-boson masses in the MSSM.  
In a first step we investigated the differences in the \order{p^2 
\alt\als} corrections as obtained in \citeres{Mh-p2-BH4} and 
\cite{Mh-p2-DDVS}.  We have shown that the difference can be attributed 
to different renormalizations of the top-quark mass.  In both 
calculations an ``on-shell'' top-quark mass was employed.  The 
evaluation in \citere{Mh-p2-BH4} includes the \order{\eps} terms of the 
top-quark mass counterterm, $\dmteps$, however, whereas this 
contribution was omitted in \citere{Mh-p2-DDVS}.  We have shown 
analytically that the terms involving $\dmteps$ do \emph{not} cancel in 
the \order{p^2\alt\als} corrections to the renormalized Higgs-boson 
self-energies (an effect that was already observed in the 
\order{\alt\als} corrections in the NMSSM Higgs 
sector~\cite{Mh-altals-NMSSM}).  
Numerical agreement between 
\citeres{Mh-p2-BH4} and \cite{Mh-p2-DDVS} is found as soon as the 
$\dmteps$ terms are dropped from the calculation in \citere{Mh-p2-BH4}.  
Moreover, as an alternative interpretation,  
we have shown that omitting the $\dmteps$ terms 
is equivalent to a redefinition of the finite part of the two-loop 
field-renormalization constant which affects the Higgs-boson mass
prediction at the three-loop order (apart from a numerically 
insignificant shift in $\tb$ as an input parameter).  
The differences between the two calculations can thus be regarded as an 
indication of the size of the missing momentum-dependent corrections 
beyond the two-loop level, and reach up to several hundred MeV in the 
case of the light $\cp$-even Higgs-boson.

In a second step we performed a calculation of the \order{\alt\als} and 
\order{p^2\alt\als} corrections employing a \DRbar\ top-quark mass 
counterterm.  We analyzed the numerical difference of the Higgs-boson 
masses evaluated with $\dmtOS$ and with $\dmtDRbar$.  By varying the 
$\cp$-odd Higgs-boson mass, $\MA$, the gluino mass, $\mgl$ and the 
off-diagonal entry in the scalar-top mass matrix, $\Xt$, we found that 
in all cases the scheme dependence, in particular of the light 
$\cp$-even Higgs-boson mass, is strongly reduced by going from the full 
one-loop result to the two-loop result including the \order{\alt\als} 
corrections.  The further inclusion of the \order{p^2\alt\als} 
contributions had a numerically small effect.  The differences found at 
the two-loop level indicate that corrections at the three-loop level and 
beyond, stemming from the top/stop sector, are expected at the level of 
$\sim 3 \gev$.  This is in agreement with existing calculations beyond 
two-loop~\cite{Mh-logresum,mhiggsFD3l}.  The possibility to use 
$\mtDRbar$ instead of $\mtOS$ has been added to the \fh\ package and 
allows an improved estimate of the size of missing corrections 
beyond the two-loop order.

%\clearpage

%%%%%%%%%%%%%%%%%%%%%%%%%%%%%%%%%%%%%%%%%%%%%%%%%%%%%%%%%%%%%%%%%%%%%%%%%%%%%%%
%%%%%%%%%%%%%%%%%%%%%%%%%%%%%%%%%%%%%%%%%%%%%%%%%%%%%%%%%%%%%%%%%%%%%%%%%%%%%%%

\bigskip
\subsection*{Acknowledgements}

We thank P.~Breitenlohner, H.~Haber, S.~Jones, M.~M\"uhlleitner, 
H.~Rzehak, P.~Slavich and G.~Weiglein, 
for helpful discussions. 
The work of S.H.\ is supported in part by CICYT (grant FPA 2013-40715-P) 
and by the Spanish MICINN's Consolider-Ingenio 2010 Program under Grant
MultiDark No.\ CSD2009-00064. S.B.\ gratefully acknowledges financial support 
by the ERC Advanced Grant MC@NNLO (340983).
This research was supported in part by the Research Executive Agency
(REA) of the European Union under the 
Grant Agreement PITN-GA2012-316704 (HiggsTools).

%%%%%%%%%%%%%%%%%%%%%%%%%%%%%%%%%%%%%%%%%%%%%%%%%%%%%%%%%%%%%%%%%%%%%%%%%%%%%%%
%%%%%%%%%%%%%%%%%%%%%%%%%%%%%%%%%%%%%%%%%%%%%%%%%%%%%%%%%%%%%%%%%%%%%%%%%%%%%%%

\end{document}